\documentclass[preprintnumbers,article,amsmath,amssymb,floatfix,10pt,prd,twocolumn,
superscriptaddress,nofootinbib]{revtex4-2}
\usepackage{bm}
\usepackage{amsfonts}
\usepackage{latexsym}
\usepackage[latin1]{inputenc}
\usepackage{graphicx}
\usepackage{amsmath}
\usepackage{palatino}
\usepackage{mathpazo}
\usepackage{textcomp}
\usepackage{float}
\usepackage{booktabs}
\usepackage{bigints} 
\usepackage{dcolumn}
\usepackage{ragged2e}
\usepackage{hyperref}
\hypersetup{colorlinks,citecolor=blue}
\hypersetup{colorlinks=true,linkcolor=red,filecolor=magenta,    urlcolor=blue}
\usepackage{amsmath}
\usepackage{xcolor}
\usepackage{orcidlink}
\usepackage{epsfig}
\usepackage{subfigure}
\usepackage{commath}
\usepackage{mathtools}

\def\jnl@style{\it}
\def\aaref@jnl#1{{\jnl@style#1}}

\def\aaref@jnl#1{{\jnl@style#1}}

\def\aj{\aaref@jnl{AJ}}                   
\def\apj{\aaref@jnl{ApJ}}                 
\def\apjl{\aaref@jnl{ApJ}}                
\def\apjs{\aaref@jnl{ApJS}}               
\def\apss{\aaref@jnl{Ap\&SS}}             
\def\aap{\aaref@jnl{A\&A}}                
\def\aapr{\aaref@jnl{A\&A~Rev.}}          
\def\aaps{\aaref@jnl{A\&AS}}              
\def\mnras{\aaref@jnl{Mon.~Not.~Roy.~Astron.~Soc.}}             
\def\prd{\aaref@jnl{Phys.~Rev.~D}}        
\def\prc{\aaref@jnl{Phys.~Rev.~C}}  
\def\prl{\aaref@jnl{Phys.~Rev.~Lett.}}    
\def\qjras{\aaref@jnl{QJRAS}}             
\def\skytel{\aaref@jnl{S\&T}}             
\def\ssr{\aaref@jnl{Space~Sci.~Rev.}}     
\def\zap{\aaref@jnl{ZAp}}                 
\def\nat{\aaref@jnl{Nature}}              
\def\aplett{\aaref@jnl{Astrophys.~Lett.}} 
\def\apspr{\aaref@jnl{Astrophys.~Space~Phys.~Res.}} 
\def\physrep{\aaref@jnl{Phys.~Rep.}}      
\def\physscr{\aaref@jnl{Phys.~Scr}}       
\def\commat{\aaref@jnl{Comm.~Math.~Phys.}}              
\def\science{\aaref@jnl{Science}}               
\def\cqg{\aaref@jnl{Classical Quant.~Grav.}}            
\def\jpcs{\aaref@jnl{JPCS}}                                     
\def\ijmpd{\aaref@jnl{Int.~J.~Mod.~Phys.~D}}                    
\def\grg{\aaref@jnl{Gen.~Relat.~Gravit.}}               
\def\rpp{\aaref@jnl{Rep.~Prog.~Phys.}}          
\def\npa{\aaref@jnl{Nucl.~Phys.~A}}        
\def\lrr{\aaref@jnl{Living Rev.~Rel.}}                   
\def\jcap{\aaref@jnl{J.~Cosmology Astropart.~Phys.}}    
\def\rmp{\aaref@jnl{Rev.~Mod.~Phys.}}   
\def\epjc{\aaref@jnl{Eur.~Phys.~J.~C}} 
\def\plb{\aaref@jnl{~Phy.~Lett.~B}} 
\def\mpla{\aaref@jnl{Mod.~Phy.~Lett.~A}} 
\def\arxiv{\aaref@jnl{arxiv.org}}

\allowdisplaybreaks[1]

\begin{document}
\color{black} 
\title{Role of cosmic voids and their matter properties in shaping wormhole geometry in generalized geometry-matter coupling gravity}


\author{A. Errehymy\orcidlink{0000-0002-0253-3578}}
\email{abdelghani.errehymy@gmail.com}
\affiliation{Astrophysics Research Centre, School of Mathematics, Statistics and Computer Science, University of KwaZulu-Natal, Private Bag X54001, Durban 4000, South Africa}
\affiliation{Center for Theoretical Physics, Khazar University, 41 Mehseti Str., Baku, AZ1096, Azerbaijan}

\author{S. Hansraj\orcidlink{0000-0002-8305-7015}}
\email{hansrajs@ukzn.ac.za}
\affiliation{Astrophysics Research Centre, School of Mathematics, Statistics and Computer Science, University of KwaZulu-Natal, Private Bag X54001, Durban 4000, South Africa}

\date{\today}
\begin{abstract}
Cosmic voids are increasingly recognized as a promising tool for cosmological exploration. Their distribution and density profiles are highly responsive to alterations in gravitational theories, along with the influences of dark energy and neutrinos. Investigating voids offers a compelling opportunity to uncover signatures of alternative gravity models on a cosmological level. Voids span a notable range of density contrasts, from approximately $-1$ near their centers to around $0$ at their edges, where screening mechanisms become less effective. The primary objective of this study is to explore a novel model that introduces a new category of wormhole solutions by leveraging cosmic voids---vast underdense regions of the universe---for the first time. We focus on identifying new exact static wormhole models by proposing an alternative viewpoint on their matter content, rooted in the recently formulated $f(R,\mathcal{L}_m,T)$ theory of gravity. By developing a unique solution based on a universal density profile for voids, we examine crucial constraints on the parameters that dictate matter distribution and the structure of spacetime itself. Our results reveal how these cosmic voids significantly influence the geometry of wormholes, steepening the gradient toward the throat and mitigating violations of the null and weak energy conditions, particularly beyond their centers. We also highlight intriguing gravitational lensing effects, showing that this wormhole repels light rather than capturing it, creating a fascinating interaction between gravity and light. Furthermore, we investigate the stability of the solution using the TOV formalism, along with the effects of exotic matter, the exoticity parameter, and anisotropy on wormhole geometry. These insights contribute to a deeper understanding of wormhole behavior in underdense environments and the pivotal role cosmic voids play in their formation.
\end{abstract}

\maketitle


\section{Introduction}\label{ch: I}
Einstein's General Relativity (GR) offers an insightful perspective on the Universe, particularly through what we call the cosmic concordance model, or $\Lambda-$CDM. This model aligns well with many observations on a cosmic scale \cite{SDSS:2005xqv, SupernovaCosmologyProject:2008ojh, Allen:2007ue, WMAP:2012nax, Planck:2018vyg}. However, one of the biggest mysteries in modern cosmology is dark energy. This strange form of energy, which has negative pressure, is necessary to explain the Universe's accelerated expansion that we've observed at lower redshifts \cite{SupernovaCosmologyProject:1997zqe, SupernovaSearchTeam:1998fmf}. The most straightforward candidate for dark energy is the cosmological constant, represented by the symbol $\Lambda$, which serves as a free parameter in Einstein's equations. 

There are several alternative theories that suggest modifying GR to explain accelerating solutions. These models propose that GR might not fully capture certain aspects of the cosmos, which leads to the introduction of new factors to better describe gravitational behavior on smaller scales (see, for example, \cite{Dolgov:2003px, Nojiri:2006ri, Clifton:2011jh, Joyce:2014kja, Ishak:2018his}). Modified gravity models, in particular, often mimic the effects of the cosmological constant in the context of the Universe's expansion. To align with the strict tests conducted in our solar system and other precise local measurements \cite{Verrier:1859his, Bertotti:2003rm, Will:2005sn}, these models need to incorporate a screening mechanism that essentially restores the predictions of standard GR in dense regions \cite{Khoury:2003aq, Hinterbichler:2010es, Brax:2013fna, Brax:2014wla}. Most viable modified gravity models show significant similarities at the background level, making their distinct features detectable mainly through their effects on how structures form in the Universe at both linear and nonlinear scales. Recently, researchers have pointed out that there are strong overlaps between the effects of some of these modified gravity models and those related to massive neutrinos \cite{He:2013qha, Motohashi:2012wc, Baldi:2013iza, Giocoli:2018gqh}. Neutrinos themselves are a mysterious part of the $\Lambda-$CDM cosmology. While the standard model of particle physics initially regarded them as massless, evidence from solar neutrino oscillations has shown that they actually do have mass \cite{Becker-Szendy:1992ory, Super-Kamiokande:1998kpq, SNO:2003bmh}.

Incorporating material terms into the gravitational action is also a viable option. This approach leads to theories like $f(R, \mathcal{L}_m)$ \cite{Harko:2010mv}, $f(R, T)$ \cite{Harko:2011kv}, and $f(R,\mathcal{L}_m,T)$ \cite{Haghani:2021fpx}. Here, $R$ stands for the Ricci scalar, $\mathcal{L}_m$ is the matter Lagrangian density, and $T$ is the trace of the energy-momentum tensor. Adding $\mathcal{L}_m$ and/or $T$ to the gravitational action expands the $f(R)$ theory. This is important because the standard $f(R)$ gravity struggles with explaining how the universe transitions from slowing down to speeding up, dealing with singularities in dark energy models \cite{Amendola:2006kh, Amendola:2006we}, and passing various tests conducted in our solar system. Additionally, relativistic stars cannot fit into the $f(R)$ framework due to the introduction of higher derivatives than 2. Moving such contributions to the matter sector of the field equations as is done customarily, negates our ability to study the simplest astrophysical phenomena such as perfect fluids.  Investigations  in torsion and non-metricity theories that also include matter fields, are well documented in the literature \cite{Junior:2015bva, Xu:2019sbp, Najera:2021afa}. Other distinct theoretical frameworks can also be found in \cite{Ilyas:2025fal, Errehymy:2025nzt, Errehymy:2025azt, Ilyas:2024ewk, Ilyas:2024dse, Errehymy:2024tqr, Errehymy:2024sme, Errehymy:2024lhl, Battista:2024gud, Ilyas:2023jhf, Ilyas:2023rde, Ilyas:2023ffm, Athar:2022lxw}.

By definition, cosmic voids are vast regions in the Universe where matter is significantly less dense. These areas have some of the lowest densities found anywhere, often much lower than the average density of the Universe, but they're not entirely empty. There are different ways to define voids; depending on the definition, some may be completely devoid of galaxies, while others simply have fewer galaxies compared to their surroundings. Voids are one of the four main components that make up the large-scale structure of the Universe, which also includes walls, filaments, and knots. Together, these components create complex, web-like structures known as the cosmic web, showcasing the intricate and fascinating arrangement of matter throughout the cosmos. The discovery of cosmic voids happened alongside the unveiling of the cosmic web. Before this revelation in the 1970s, people knew about galaxies and galaxy clusters, and they understood that the Universe was expanding, but the large-scale distribution of these structures was largely a mystery. In 1978, two independent studies using different galaxy surveys both confirmed the existence of intergalactic voids. J\^{o}eveer et al. \cite{Joeveer:1978} and Gregory and Thompson \cite{Gregory:1978} found something surprising: the voids were much larger than expected, ranging from tens to hundreds of megaparsecs. They pointed out that there are vast regions of space, with radii greater than 20 megaparsecs, where there appeared to be no galaxies at all. This provided the first observational evidence for cosmic voids and demonstrated their connection to the cosmic web.

The density found within cosmic voids is usually about 10\% of the Universe's average density. Interestingly, these voids can also contain smaller structures known as sub-cosmic webs, which have noticeable fluctuations in density \cite{Aragon-Calvo:2012ewn, Jaber:2023rjx}. The size of a void can vary, with effective radii ranging from just a few megaparsecs to around 100 megaparsecs. In fact, voids fill a significant portion of the Universe's volume. According to various algorithms designed to identify voids, they make up over 60\% of the total volume in the later stages of the Universe \cite{Cautun:2014fwa, Libeskind:2017tun}. Additionally, matter surrounding the centers of these voids tends to drift away from them, and the edges of a typical void expand at a rate faster than the overall expansion of the Universe. The statistical features of cosmic voids are shaped by the Universe's initial conditions. In the standard $\Lambda-$CDM model, these conditions are influenced by specific cosmological parameters. Unlike other regions of the Universe, cosmic voids are more directly affected by cosmic expansion and the overall dynamics at play. As a result, they provide valuable insights into both cosmology and astrophysics. For instance, by examining the density profiles of these voids and their volume distribution, scientists can estimate the total mass of neutrinos, test different theories of gravity, and refine our understanding of cosmological parameters. Cosmic voids are increasingly recognized as important tools for understanding cosmology. Research based on galaxy redshift surveys has uncovered effects like redshift-space distortions \cite{Kaiser:1987qv} and the Alcock-Paczynski effect \cite{Alcock:1979mp} related to these voids. By adding more data, such as the cosmic microwave background (CMB) from shortly after the Big Bang and observations of background galaxies, scientists have also been able to detect phenomena like gravitational lensing and the integrated Sachs-Wolfe effect \cite{Sachs:1967er}. These discoveries help deepen our understanding of how matter clusters in the Universe, how matter perturbations grow, and the overall makeup of matter and energy in the cosmos.

Wormholes represent an intriguing concept in Einstein's field equations, functioning as connections between two distinct universes or remote regions within the same universe. However, the first wormhole solutions studied by Flamm \cite{Flamm:1916plb} were found to be inherently unstable, which means they wouldn't allow for safe travel. The Einstein-Rosen bridge, while detailed mathematically by Einstein and Rosen \cite{Einstein:1935tc}, isn't a traversable wormhole--people can't actually travel through it. The concept of a ``wormhole" was introduced by Misner and Wheeler in 1957 \cite{Misner:1957mt}, and some of these theoretical wormholes might be large enough for humans to use and could even make time travel possible \cite{Morris:1988cz}. A true traversable wormhole would ideally not have horizons or singularities, making it a more promising avenue for exploration. A significant breakthrough occurred when Ellis \cite{Ellis:1973yv} discovered a unique wormhole solution based on a spherically symmetric arrangement of Einstein's equations, which included a massless scalar field with unusual properties. A key aspect of wormhole formation in GR is that it often requires violating certain energy conditions \cite{Visser:1995yv}. This can lead to skepticism about whether wormholes could actually exist. As a result, matter distributions that go against the null energy condition are labeled as exotic and are often seen as having limited physical relevance, mainly because there's not much experimental evidence to support their existence. It has been increasingly recognized that wormhole spacetimes can emulate several key characteristics traditionally associated with black holes. For instance, Damour and collaborators \cite{Damour:2007ap} pointed out that phenomena such as quasi-normal mode oscillations, accretion dynamics, and even aspects related to the no-hair conjecture---long believed to be hallmarks of black holes---can also emerge in the context of wormhole geometries. Their analysis revealed that wormholes can, under certain conditions, closely mimic the external appearance and behavior of black holes, making them nearly indistinguishable observationally. Building on this, Cardoso et al. \cite{Cardoso:2016rao} presented compelling evidence that a subclass of wormholes---specifically those supported by a thin shell of exotic (phantom) matter at the throat---can produce quasi-normal mode signals nearly identical to those of black holes in the early stages of gravitational wave emission. These findings were particularly timely, coinciding with the first direct detection of gravitational waves by LIGO \cite{LIGOScientific:2016aoc}, and raised important questions about the true nature of the compact objects responsible for such signals. According to their work, deviations from the standard black hole ringdown signature become noticeable only at late times, thereby allowing wormholes to remain observationally hidden behind a black hole-like appearance in initial detections. Further investigations by Konoplya et al. \cite{Konoplya:2016hmd} broadened this perspective by demonstrating that wormholes can exhibit a range of ringing behaviors---some that strongly resemble black holes and others that diverge significantly---depending on the underlying geometry and matter content. These studies collectively highlight the need for more refined observational strategies to differentiate between black hole and wormhole candidates. Given the growing body of research suggesting that wormholes could serve as viable black hole mimickers, it is both necessary and exciting to pursue further theoretical and observational studies aimed at identifying distinctive signatures. For a broader overview of ongoing efforts in this area, including those that examine lensing, shadow formation, and accretion disk imaging around compact objects, we refer readers to Refs.~\cite{Bambi:2013nla, Nedkova:2013msa, Ohgami:2015nra, Shaikh:2017zfl, Jusufi:2017mav,Shaikh:2018kfv,Shaikh:2018oul,Dai:2019mse}. Additional insights into accretion disk images and their potential to distinguish compact object geometries can be found in \cite{Luminet:1979nyg,Stuchlik:2010zz,Stuchlik:2014yaa,Schee:2015nua,Gyulchev:2019tvk,Tian:2019yhn}. Recently \cite{ban-hans-prad-abd} devised a wormhole model where the Casimir stresses evident in the equation of state was used as a physical mechanism to provide a repulsive gravitational effect to keep the wormhole throat open sufficiently to be traversable.

Wormholes remain elusive, making our primary objective in studying these intriguing solutions the enhancement of our theoretical knowledge. As more researchers publish papers examining various aspects of wormholes, we strive to clarify their physical characteristics. This journey often requires us to make different assumptions about the matter that constitutes these structures, some of which can simplify the complex equations we face. Fortunately, several compelling exact wormhole models have already emerged from general relativity and certain modified gravity theories. However, it has been shown that cosmic voids---vast, low-density regions scattered throughout the Universe---aren't just empty spaces; they follow surprisingly consistent density patterns that rise sharply from their centers out toward the denser cosmic web \cite{Sheth:2003py, Hamaus:2014fma}. This steep inner profile offers more than just a cosmological curiosity---it provides useful intuition for building physically plausible wormhole models. If a wormhole throat sits within a void-like environment, the rapidly increasing density near the center can help balance or soften the exotic matter usually needed to keep the throat open \cite{Kuhfittig:2013hva, Lobo:2005us}. In effect, the void structure can mask or reduce the violation of energy conditions, making the wormhole spacetime behave more like standard matter in some regions. Recent studies even suggest that voids could play a key role in testing alternative gravity theories like  $f(R)$, where matter and geometry are closely linked \cite{Cai:2014fma, Pisani:2019cvo}. By drawing from how real cosmic structures behave, we can anchor wormhole models in more realistic, observable physics.

In this paper, we will investigate a new model that introduces a novel class of wormhole solutions that have not been previously documented. Our emphasis will be on discovering new exact static wormhole models by proposing an alternative perspective on their matter content, grounded in the recently developed $f(R,\mathcal{L}_m,T)$ theory of gravity. This theory has already proven itself in various exciting applications, such as analyzing thermodynamics and perturbations of $f(R,\mathcal{L}_m,T)$ models \cite{Zubair:2023umm}, as well as exploring the configurations of compact objects like compact stars \cite{Mota:2024kjb, Fortunato:2024ulg} and surrounding wormhole solutions \cite{Moraes:2024mxk, Errehymy:2025kzj}. Furthermore, the original authors of this theory \cite{Haghani:2021fpx} have compared its predictions with cosmological data, enriching our understanding of the topic. Consequently, we aim to construct an exact wormhole model by assuming that the energy density of the matter within the wormhole can be described in terms of cosmic voids. Investigating cosmic voids within modified gravity theories is crucial for enhancing our understanding of how gravity operates in unconventional scenarios. This approach not only deepens our insight into the fundamental nature of spacetime but also enables us to tackle significant questions about gravity and its interactions with other fundamental forces. By integrating cosmic voids into our research, we can uncover valuable insights that transcend the limitations of traditional gravity theories and pave the way for new explorations.

This paper is structured to guide readers through the findings step by step. In Sec. \ref{ch: II}, the discussion begins with the history and unique characteristics of cosmic voids, along with what defines a traversable wormhole. Moving on to Sec. \ref{ch: III}, the focus shifts to the field equations for spherically symmetric wormholes within the framework of $f(R,\mathcal{L}_m,T)$ gravity. In Sec. \ref{ch: IV}, specific solutions related to cosmic voids and the energy conditions associated with local sources of matter are explored. Sec. \ref{ch: V} highlights the implications of the theory through gravitational lensing effects. Next, in Sec. \ref{ch: VI}, a stability analysis is conducted using the TOV equation. Sec. \ref{ch: VII} investigates how exotic matter, the exoticity parameter, and anisotropy influence wormhole geometry. Finally, in Sec. \ref{ch: VIII}, the paper concludes with a summary of key findings and a thoughtful discussion on their implications for understanding the subject.

\section{Cosmic voids}\label{ch: II}
Cosmic voids are vast, underpopulated regions in the universe. First identified by Gregory and Thompson in 1978 \cite{Gregory:1978}, and later studied by Kirshner and others \cite{Kirshner:1981,de Lapparent:1986} in the early 1980s, these voids have become essential for our understanding of the cosmos \cite{Weygaert:2016, Pisani:2019}. They constitute a significant portion of the universe's volume \cite{Ceccarelli:2013}, making them the largest observable structures. As such, cosmic voids provide valuable insights into the universe's expansion and overall shape. The formation of large-scale underdensities is a natural aspect of cosmic structure development. As the universe evolves, galaxies are drawn toward areas with more matter due to gravitational forces. This movement leads to the creation of various structures, such as groups, clusters, and filaments. In the process, galaxies leave behind underdense regions that become increasingly empty, giving rise to cosmic voids. This relationship illustrates how voids and other structures are two sides of the same coin in cosmic evolution. Moreover, cosmic voids are connected to the tiny density fluctuations from the early universe, which are reflected in the slight variations observed in the Cosmic Microwave Background (CMB) \cite{Cai:2014, Cai:2017}. Understanding the statistical properties of voids is influenced by two key factors. First, the choice of matter tracers for mapping large-scale structures---such as galaxies, dark-matter halos, and dark-matter particles---is crucial. Second, the methods used to identify these voids based on the distribution of these tracers are equally important. Various void-finding algorithms exist, and for a comparison of these techniques, \cite{Colberg:2008} provides valuable insights. When investigating the properties of voids, it's essential to consider how they are identified. For instance, researchers like Ceccarelli \cite{Ceccarelli:2013} and Paz \cite{Paz:2013} employed a spherical void finder in their studies. In contrast, Hamaus et al. \cite{Hamaus:2014fma} introduced a cosmic void based on a universal density profile that does not differentiate between void types. They achieved this using a modified version of the ZOnes Bordering On Voidness void finder, which utilizes the watershed algorithm:
\begin{eqnarray}\label{RhoCV}
\rho(r) = \rho_a \left( 1 + \delta_c \frac{1 - \left( \frac{r}{r_{sc}} \right)^\alpha }{1 + \left( \frac{r}{r_{sv}} \right)^\beta }\right),
\end{eqnarray}
where $r_{sc}$ represents the scale at which the density of the void matches the average density of its surroundings, and $r_{sv}$ denotes the size of the void. The parameter $\delta_c$ (which is negative) quantifies the contrast, reflecting the emptiness of the void---specifically, $\delta_c = -1$ indicates a perfect vacuum at $r = 0$. Here, $\rho_a$ is the average density of the universe, and $\alpha$ and $\beta$ are empirical fitting parameters. In this universal void density profile each parameter carries specific physical significance and constraints. The central density contrast $\delta_c$ lies in the range $-1 < \delta_c < 0$, with $\delta_c = -1$ representing a perfect vacuum at the center, and values closer to 0 indicating shallower voids. The compensation scale $r_{sc}$ is the radius where the void's density equals the cosmic mean, constrained to $0 < r_{sc} < r_{sv}$. The void radius $r_{sv}$ defines the outer boundary of the void and typically ranges from $\sim 5$ to $100\ h^{-1}$ Mpc in observations. The parameter $\alpha > 0$ controls the steepness of the density rise from the void center to $r_{sc}$, with typical values in simulations around $1$ to $2$. The outer slope parameter $\beta$ governs the falloff beyond the void edge, often taken in the range $1 \lesssim \beta \lesssim 10$, with higher values indicating sharper transitions. These parameters are usually determined through fits to simulations or galaxy surveys, and their interpretation depends on the choice of tracers and void-finding methodology. The metric presented here captures the essence of the most general form of a static, spherically symmetric spacetime, as discussed in \cite{II: CAR}:
\begin{align} \label{1}
ds^{2}=-e^{\tilde{\nu}(r)}dt^{2}+e^{\tilde{\lambda}(r)} dr^{2}+r^{2} d\theta^{2}+r^{2}\sin^{2}\theta d\phi^{2}.
\end{align}
In this analysis, we will introduce several important variables. The symbol $t$ stands for time and can take on any real value. The variable $r$ represents the radial coordinate, starting at a minimum value $r_0$ and stretching out to infinity. The angles $\theta$ and $\phi$ define the spherical coordinates, with $\theta$ ranging from 0 to $\pi$ and $\phi$ going from 0 to $2\pi$. We assume that the functions $\tilde{\nu}(r)$ and $\tilde{\nu}(r)$ are smooth, meaning they change gradually as $r$ varies, without any sudden jumps. For our purposes, we will define $\tilde{\nu}(r) = 2\Phi(r)$, where $\Phi(r)$ is known as the \textit{redshift} function. This function plays a vital role in calculating gravitational redshift, which we will derive using the metric from Eq. \eqref{1}. In the framework of $f(R, \mathcal{L}_m, T)$ gravity, assuming a \textit{constant redshift function}, $\Phi(r) = \text{const}$, is a strategic and physically motivated simplification when constructing traversable wormhole solutions. This assumption ensures that \textit{no event horizon} forms within the geometry, which is crucial for ensuring that signals or particles can travel through the wormhole without encountering infinite redshift. If $\Phi(r)$ were to diverge, the resulting time dilation would effectively block traversal. Additionally, since $f(R, \mathcal{L}_m, T)$ gravity introduces highly nonlinear field equations---owing to the interplay between the Ricci scalar, matter Lagrangian $\mathcal{L}_m$, and the trace of the energy-momentum tensor $T$---keeping $\Phi(r)$ constant removes higher-order derivative terms and significantly simplifies the mathematics. This makes it more practical to extract exact or approximate solutions for the shape function and matter profile. Physically, a constant $\Phi(r)$ implies that the gravitational potential is radially uniform, resulting in a consistent time rate throughout the wormhole, which aids in smoothly connecting interior wormhole regions to an external spacetime. More importantly, since this class of modified gravity can inherently violate energy conditions without needing exotic matter, a constant redshift helps isolate and clarify the role of the modified gravitational dynamics in maintaining the wormhole structure. However, if one considers a \textit{nonconstant redshift function}, the picture becomes richer but more complicated. A variable $\Phi(r)$ introduces additional terms into the gravitational field equations---specifically $\Phi'(r)$ and $\Phi''(r)$---which influence the balance between geometry and matter content. These extra terms can shift the dynamical stability of the configuration, especially when perturbations are introduced. Depending on how $\Phi(r)$ varies, the wormhole could become more or less stable under small fluctuations. From the perspective of \textit{gravitational lensing}, allowing $\Phi(r)$ to vary affects the bending of light near the throat, as the redshift function contributes to the effective optical geometry. While a constant redshift often produces deflection angles that hint at a repulsive gravitational signature---characteristic of many wormholes---a variable $\Phi(r)$ could either strengthen, weaken, or even reverse this behavior depending on its radial profile. Thus, although the constant redshift assumption offers analytical convenience and highlights the role of modified gravity in wormhole physics, relaxing this condition opens up a broader---and potentially more realistic---landscape of wormhole solutions with diverse stability and observational signatures. The $g_{tt}$ component of the metric is crucial for defining energy, making $\Phi$ essential in understanding the redshift effect. Additionally, we will follow a common approach found in the literature on wormholes by expressing $e^{\tilde{\nu}(r)}$ as $\left(1 - \frac{\hat{X}(r)}{r}\right)^{-1}$. This choice is based on the work of \cite{Morris:1988cz}. In this context, $\hat{X}(r) \equiv \hat{X}$ is referred to as the \textit{shape function}. To ensure that a wormhole can be traversed, certain conditions need to be met, as shown in \cite{Morris:1988cz}:
\begin{align}\label{eq: TWC}
 \begin{split}
        \frac{\hat{X}}{r}\!<\!1\,\,\forall r\!>\!r_0;\,\hat{X}(r_0)\!=\!r_0\quad&\textit{Throat Condition}\\
        \frac{\hat{X}-\hat{X}'r}{\hat{X}^2}\biggr\rvert_{r_0}\!\!>\!0\rightarrow \hat{X}'(r_0)\!<\!1, \quad&\textit{Flaring-Out Condition}\\
        \lim_{r\rightarrow\infty}\frac{\hat{X}}{r}=0\quad&\textit{Asymptotic Flatness Condition}.
    \end{split}   
\end{align} 
To prevent horizons from forming, it's really important that the redshift function stays bounded for all values of $r$. Luckily, the choice of $\Phi(r)$ that we'll look at shortly will take care of this without any problems. The metric we described in Eq. \eqref{1} needs to align with the Einstein field equations. By picking the right energy-momentum tensor, we can figure out the functions $\hat{X}$ and $\Phi$. However, as Morris and Thorne pointed out \cite{Morris:1988cz}, if we want to find solutions for traversable wormholes, we need to include something called \textit{exotic} matter-types of matter that break certain energy conditions. To work around this issue, we can tweak the standard Einstein equations a bit. This method, which focuses on curvature as a key idea, is known as $f(R)$ gravity. That said, this approach does have some drawbacks: either the wormholes can't be asymptotically flat, or we might find that the effective gravitational constant changes sign in certain regions of spacetime \cite{I: BSS}. This second issue is particularly important when we look at lensing effects, which we'll discuss in Sec. \ref{ch: V}.
 
\section{Field equations for wormholes with spherically symmetric in $f(R,\mathcal{L}_m,T)$ gravity}
\label{ch: III}
Let's start by discussing the gravitational action functional, which we express in units where $G=c=1$, following the approach from \cite{I: HAG}:
\begin{equation}\label{eq: Ac}
\mathcal{S}=\frac{1}{16\pi}\int\,f(R,\mathcal{L}_m,T)\sqrt{-g}\,d^4x+\int \mathcal{L}_m\,\sqrt{-g}\,d^4x\,.
\end{equation}
In this formulation, the Ricci scalar $R$ acts as a measure of curvature and is closely related to the metric tensor $g_{\mu\nu}$, whose determinant is denoted as $g$. To better understand how $R$ connects with $g_{\mu\nu}$, we can look at the metric-affine connection, which does not involve torsion:
\begin{equation}
    \Gamma^\alpha_{\beta\gamma}= \frac{1}{2} g^{\alpha\lambda} \left( \frac{\partial g_{\gamma\lambda}}{\partial x^\beta} + \frac{\partial g_{\lambda\beta}}{\partial x^\gamma} - \frac{\partial g_{\beta\gamma}}{\partial x^\lambda} \right).
    \label{eq: CHR}
\end{equation}
Now, let's introduce our framework, which we call $f(R,\mathcal{L}_m, T)$ gravity. This framework simplifies the more complex theory of $f(R,\mathcal{L}_m, T)$ proposed by \cite{I: HAG}. In our model, we establish a connection to the trace of the energy-momentum tensor $T$ and make a linear adjustment to the matter Lagrangian $\mathcal{L}_m$. Additionally, we replace the conventional matter coupling constant with a more flexible option. We start by understanding how to define a covariant derivative, which reflects the characteristics of curved geometry. This foundation allows us to construct the Ricci tensor, given by:
\begin{equation}\label{eq: RIC}
R_{\mu\nu} = \partial_\lambda \Gamma^\lambda_{\mu\nu} - \partial_\nu \Gamma^\lambda_{\lambda\mu} + \Gamma^\sigma_{\mu\nu} \Gamma^\lambda_{\sigma\lambda} - \Gamma^\lambda_{\nu\sigma} \Gamma^\sigma_{\mu\lambda}\,.
\end{equation}
When we contract the Ricci tensor, we derive what's known as the Ricci scalar:
\begin{equation}\label{eq: RICS}
R = g^{\mu\nu} R_{\mu\nu}\,,
\end{equation}
which provides a measure of curvature that remains consistent regardless of the coordinate system we use. This implies that we've created an action functional that is invariant under the symmetry transformations of spacetime (known as diffeomorphisms), as long as the matter Lagrangian does not depend on the coordinates. By varying the action $\mathcal{S}$ with respect to the metric tensor $g_{\mu\nu}$ and ensuring the action remains stationary, we can derive our modified Einstein equations:
\begin{multline}\label{6}
f_{R}R_{\mu \nu } - \frac{1}{2}\left[ f - (f_{\mathcal{L}_m} + 2f_{T})\mathcal{L}_m \right] g_{\mu \nu } + \left( g_{\mu \nu } \Box - \nabla_{\mu} \nabla_{\nu} \right) f_{R} \\
= \left[ 8\pi + \frac{1}{2}(f_{\mathcal{L}_m} + 2f_{T}) \right] T_{\mu \nu } + f_{T} \tau_{\mu \nu },
\end{multline}
where $f_{R} = \frac{\partial f(R,\mathcal{L}_m,T)}{\partial R}$, $f_{\mathcal{L}_m} = \frac{\partial f(R,\mathcal{L}_m,T)}{\partial \mathcal{L}_m}$, $f_{T} = \frac{\partial f(R,\mathcal{L}_m,T)}{\partial T}$, and $\tau_{\mu \nu} = 2g^{\alpha \beta} \frac{\partial^2 \mathcal{L}_m}{\partial g^{\mu\nu} \partial g^{\alpha\beta}}$. The energy-momentum tensor for the cosmic fluid, denoted by $T_{\mu\nu}$, is expressed as:
\begin{equation}
T_{\mu\nu} = \frac{-2}{\sqrt{-g}} \frac{\delta(\sqrt{-g}\mathcal{L}_m)}{\delta g^{\mu\nu}}.
\end{equation}
We assume this fluid is anisotropic, characterized by the following diagonal form:
\begin{equation}
\label{EMT}
T^\mu_{\,\,\,\,\nu} = \text{diag}(-\rho, P_r, P_t, P_t),
\end{equation}
which leads us to the expression (following \cite{Morris:1988cz, I: VIS, I: KUH}):
\begin{equation}
T_{\mu\nu} = \left(\rho + P_t\right) u_{\mu} u_{\nu} + P_t \delta_{\mu\nu} + \left(P_r - P_t\right) v_{\mu} v_{\nu}\,.
\end{equation}
In this context, $u_\mu$ represents the four-velocity, while $v_\mu$ is a unit spacelike covector, both normalized to $\mp 1$. Here, $\rho$ indicates the energy density, with the radial and tangential pressures, $P_r$ and $P_t$, depending on the radial coordinate $r$. As a result, the modified Einstein equations \eqref{6} can be expressed in this new framework:
\begin{widetext}
\begin{align}\label{11}
&\frac{1}{2} \rho \left(16 \pi +  f_{\mathcal{L}_m}+2f_{T}\right)+ f_{T} \mathcal{L}_m\,- \frac{1}{2} \left( f- f_{\mathcal{L}_m} {\mathcal{L}_m} \right)=\nonumber\\&\hspace{3cm}\left( 1-\frac{\hat{X}}{r} \right) \left[ \left\lbrace  \Phi''+\Phi'^2   - \frac{(r\hat{X}'-\hat{X})}{2r(r-\hat{X})}\Phi' + \frac{2\Phi'}{r}\right\rbrace f_R
\right. \left.
-\left\lbrace \Phi' - \frac{(r\hat{X}'-\hat{X})}{2r(r-\hat{X})}  +\frac{2}{r}\right\rbrace f_R' - f_R'' \right],
\end{align}
\begin{align}\label{12}
& \frac{1}{2} P_r \left(16\pi+f_{\mathcal{L}_m}+2f_{T}\right)- f_{T} \mathcal{L}_m\,+ \frac{1}{2} \left( f- f_{{\mathcal{L}_m}} {\mathcal{L}_m} \right)=\nonumber\\&\hspace{3cm}\left( 1-\frac{\hat{X}}{r} \right) \left[ \left\lbrace  - \left(\Phi''+\Phi'^2\right)  + \frac{(r\hat{X}'-\hat{X})}{2r(r-\hat{X})} \left( \Phi' +\frac{2}{r} \right) \right\rbrace f_R  
\right. \left.
+ \left\lbrace \Phi'- \frac{(r\hat{X}'-\hat{X})}{2r(r-\hat{X})}  +\frac{2}{r}\right\rbrace f_R' \right] ,
\end{align}
\begin{align}\label{13}
& \frac{1}{2} P_t \left(16\pi+f_{\mathcal{L}_m}+2f_{T}\right)- f_{T} \mathcal{L}_m\,+  \frac{1}{2} \left( f-{\mathcal{L}_m} f_{{\mathcal{L}_m}} \right)=\nonumber\\&\hspace{5cm}\left( 1-\frac{\hat{X}}{r} \right) \left[ \left\lbrace - \frac{\Phi'}{r}  + \frac{(r\hat{X}'+\hat{X})}{2r^2(r-\hat{X})} \right\rbrace f_R+ \left\lbrace \Phi'+\frac{2}{r} \right.\right. \left.\left.
- \frac{(r\hat{X}'-\hat{X})}{2r(r-\hat{X})}  \right\rbrace f_R' + f_R'' \right].
\end{align}
\end{widetext}
To investigate these equations, we will use a straightforward linear model defined as $ f(R,\mathcal{L}_m, T) = R + \eta \mathcal{L}_m + \chi T$. In this model, the parameters $\eta$ and $\chi$ indicate how much we are altering our original framework. It's important to note that when both $\eta$ and $\chi$ are set to zero, as shown in Eq. \eqref{eq: Ac}, our model simplifies back to what we know as standard GR.

To derive analytical solutions for the Eqs. \eqref{11}-\eqref{13}, we will assume that the redshift is constant, which we will denote simply as $\Phi(r) \equiv \Phi$. Previous studies have already uncovered solutions for traversable wormholes with constant redshift within the contexts of $f(R)$ gravity \cite{II: GOD, II: GOD2} and $f(R,\mathcal{L}_m)$ gravity \cite{Sol, Lakhan, Maurya:2024jos, Errehymy:2024cgy}. Next, we will calculate the Ricci scalar for the metric described in Eq. \eqref{1}:
\begin{multline} \label{15}
R=\frac{2\hat{X}'}{r^2} - 2 \left\lbrace \Phi''+\Phi'^2+\frac{\Phi'}{r} \right\rbrace \left( 1-\frac{\hat{X}}{r} \right) \\
+ \frac{\Phi'}{r^2} \left( r\hat{X}'+\hat{X}-2r \right).
\end{multline}
By setting $\mathcal{L}_m = -\rho$, as shown in the work of \cite{I: HAG}, we arrive at the following relationships:
\begin{align}\label{14a}
    \frac{\hat{X}'}{r^2} &= \rho \left(\kappa - \frac{\chi}{2}\right) - \frac{\chi}{2} (P_r + 2 P_t)\,,\\\label{14b}
    -\frac{\hat{X}}{r^3} &= P_r \left(\kappa + \frac{\chi}{2}\right) + \frac{\chi}{2} (2 P_t + \rho)\,,\\\label{14c}
    \frac{1}{2 r^2}\left(\frac{\hat{X}}{r} - \hat{X}'\right) &= P_t (\kappa + \chi) + \frac{\chi}{2} (P_r + \rho)\,,
\end{align}
where $\kappa$ is defined as $8 \pi + \frac{\eta}{2} + \chi$.
We can easily express the quantities $\rho$, $P_r$, and $P_t$ in terms of $\hat{X}$. For instance, by combining Eqs. \eqref{14a} and \eqref{14b}, we can derive an expression for the sum $(\rho + P_r)$. This result allows us to determine $P_t$ from Eq. \eqref{14c}. As a result, our system of equations \eqref{14a}-\eqref{14c} simplifies to the following forms:
\begin{align}\label{15a}
   \hat{X}' &= \kappa r^2 \rho\,,\\\label{15b}
   -\hat{X} &= \kappa r^3 P_r\,,\\\label{15c}
   \hat{X} - \hat{X}' r &= 2 \kappa r^3 P_t \,.
\end{align}
When we compare our findings with those from Moraes et al. \cite{Moraes:2024mxk}, we find that our final expressions match perfectly. This consistency adds strong support to our conclusions.

\begin{figure*}
\centering
\includegraphics[width=7.cm,height=5.75cm]{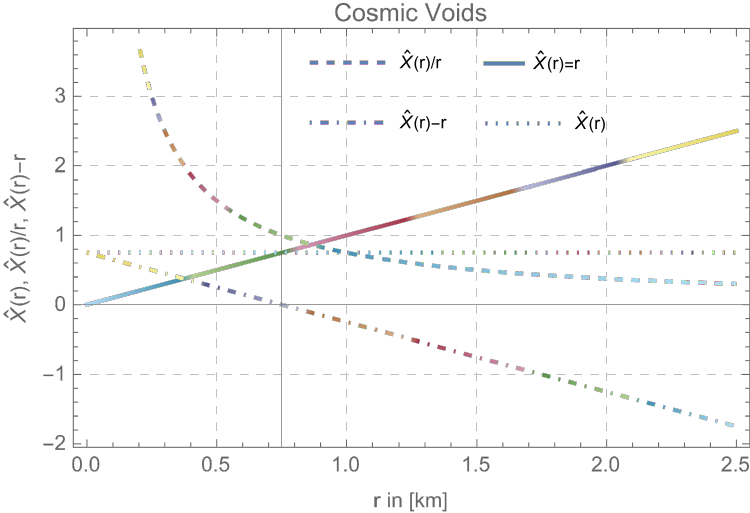}
\caption{The shape function for cosmic voids wormholes is characterized by a constant throat radius of $r_0 = 0.75$. In addition, several key parameters play a significant role: $\delta_c = -0.9$, $\lambda = 0.75$, $\chi = 0.75$, $\alpha = 3.75$, $\beta = 6.5$, $\rho_{a} = 0.00001$, $r_{Sc} = 75$ and $r_{Sv} = 95$. }\label{fig1a}
\end{figure*}
\begin{figure*}
\centering
\includegraphics[width=6.3cm,height=5.75cm]{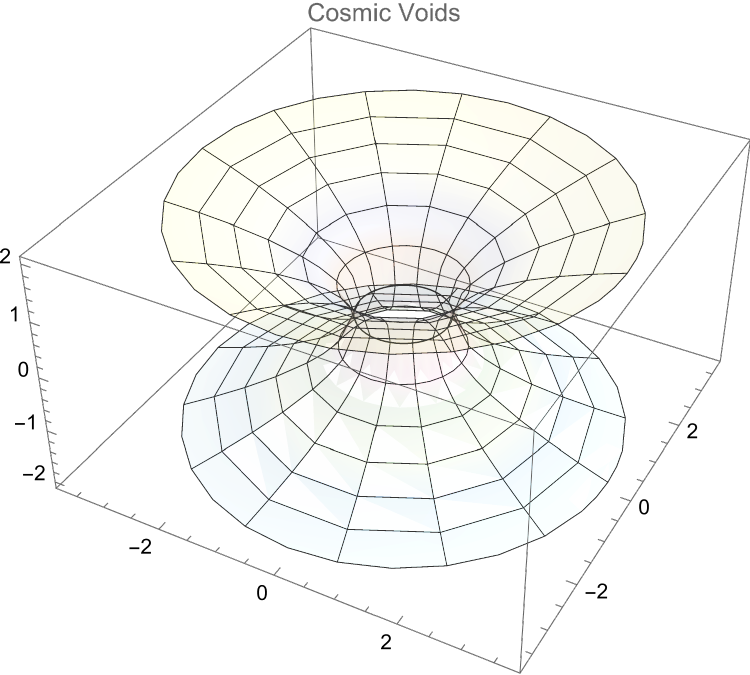}
\includegraphics[width=6.3cm,height=5.75cm]{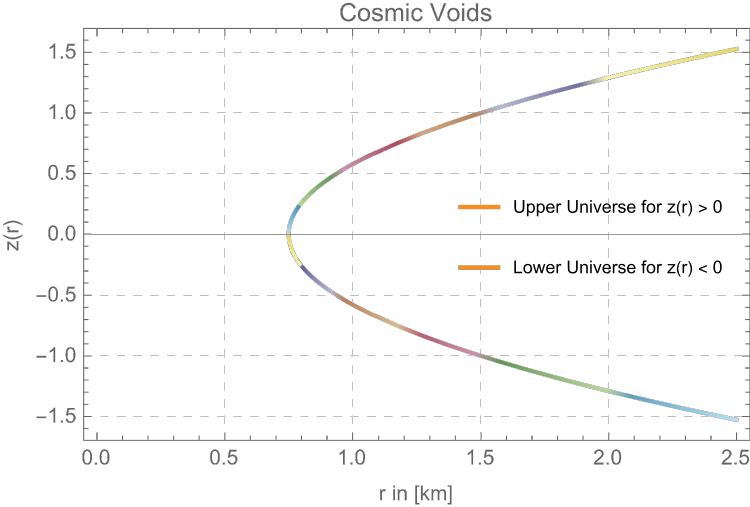}
\caption{The spacetime-embedding diagrams for cosmic voids wormholes are characterized by a constant throat radius of $r_0 = 0.75$. In addition, several key parameters play a significant role: $\delta_c = -0.9$, $\lambda = 0.75$, $\chi = 0.75$, $\alpha = 3.75$, $\beta = 6.5$, $\rho_{a} = 0.00001$, $r_{Sc} = 75$ and $r_{Sv} = 95$. }\label{fig1b}
\end{figure*}

\section{Specific solutions related to cosmic voids}
\label{ch: IV}
In this section, we turn our attention to the captivating concept of wormholes and how they relate to cosmic voids. By exploring the negative parameter $\delta_c$, which highlights the emptiness of these voids, we can better understand how this emptiness influences the nature of wormholes. To get started, we'll compare Eq. \eqref{15a} with the density profile we've chosen from Eq. \eqref{RhoCV}. This comparison is crucial because it sheds light on how cosmic voids can affect the shape and stability of a wormhole. By examining these equations together, we aim to reveal the intriguing links between energy density and the unique properties of wormholes, ultimately deepening our appreciation of these remarkable cosmic phenomena.
\begin{equation}\label{23}
 \frac{\hat{X}'}{\kappa r^2}=\rho_a \left( 1 + \delta_c \frac{1 - \left( \frac{r}{r_{sc}} \right)^\alpha }{1 + \left( \frac{r}{r_{sv}} \right)^\beta }\right).
\end{equation}
Now, let's move on to integrating the shape function. To make things simpler, we'll use the throat condition $\hat{X}(r_0) = r_0$. This handy condition helps us get rid of the integration constant, bringing us to the following expression:
\begin{widetext}
 \begin{small}
\begin{align}\label{sf}
 \hat{X}&=\frac{1}{3 (\alpha+3)}\Biggl[\Lambda r^3 \rho_a \Biggl(-3 \delta_{c} \left(\frac{r}{r_{Sc}}\right)^{\alpha} \, _2F_1\left(1,\frac{\alpha+3}{\beta};\frac{\alpha+\beta+3}{\beta};-\left(\frac{r}{r_{Sv}}\right)^{\beta}\right)+(\alpha+3) \delta_{c} \, _2F_1\left(1,\frac{3}{\beta};\frac{\beta+3}{\beta};-\left(\frac{r}{r_{Sv}}\right)^{\beta}\right)+\alpha+3\Biggl)\nonumber\\&-\Lambda r_0^3 \rho_a \Biggl(-3 \delta_{c} \left(\frac{r_0}{r_{Sc}}\right)^{\alpha} \, _2F_1\left(1,\frac{\alpha+3}{\beta};\frac{\alpha+\beta+3}{\beta};-\left(\frac{r_0}{r_{Sv}}\right)^{\beta}\right)+(\alpha+3) \delta_{c} \, \nonumber\\& _2F_1\left(1,\frac{3}{\beta};\frac{\beta+3}{\beta};-\left(\frac{r_0}{r_{Sv}}\right)^{\beta}\right)+\alpha+3\Biggl)+3 (\alpha+3) r_0\Biggl],~~~~~~
\end{align}
 \end{small}
\end{widetext}
with $_2F_1(a,b;c;z)$ being the hypergeometric function.

We can say with confidence that the new shape function, inspired by the profile of cosmic voids, effectively captures the essence of wormholes. Let's take a closer look at Fig. \ref{fig1a}, where we've plotted functions like $\hat{X}$, $\hat{X}/r$, and $\hat{X} - r$. For this analysis, we carefully selected specific parameter values: $\delta_c = -0.9, \quad \lambda = 0.75, \quad \chi = 0.75, \quad \alpha = 3.75, \quad \beta = 6.5, \quad \rho_a = 0.00001, \quad r_{Sc} = 75, \quad r_{Sv} = 95, \quad r_0 = 0.75$. As we examine Fig. \ref{fig1a}, an interesting pattern emerges. When $r$ exceeds $r_0$, the value of $\hat{X} - r$ turns negative, indicating that $\hat{X}/r < 1$. In simpler terms, as $r$ increases, the difference between $\hat{X}$ and $r$ shrinks. This aligns with the flaring-out condition for $r \geq r_0$, keeping $\hat{X}'$ below 1. Additionally, both conditions $\hat{X}' < 1$ and $\hat{X}/r < 1$ hold for all $r$ greater than $r_0$. We also observe that $\hat{X}/r$ is a decreasing function and approaches zero as $r$ tends to infinity. This consistent behavior suggests that the spacetime becomes asymptotically flat at large distances.

The embedding surface $Z(r)$ is defined according to the work by Morris and colleagues \cite{Morris:1988cz} from 1988:
\begin{equation}
      Z(r)=\pm\bigintss_{\,r_0}^{\infty}\frac{1}{\sqrt{\frac{r}{\hat{X}}-1}}\text{d}r.
 \end{equation}
 The different signs indicate the two separate branches of spacetime that are connected through the wormhole. In Fig. \ref{fig1b}, we can see spacetime-embedding diagrams that show how mass is distributed in cosmic voids. These visuals give us a fascinating look at the geometric structure of the wormhole. Now, we will rigorously derive the expressions for the radial and transverse pressure components. It's actually quite simple--just plug Eq. (\ref{sf}) into Eqs. (\ref{15b}) and (\ref{15c}):
\begin{widetext}
\begin{small}
\begin{align}
    P_r(r)&= -\frac{1}{3 (\alpha+3) \Lambda r^3}\Biggl[\Lambda r^3 \rho_{a} \Biggl(-3 \delta_{c} \left(\frac{r}{r_{Sc}}\right)^{\alpha} \, _2F_1\left(1,\frac{\alpha+3}{\beta};\frac{\alpha+\beta+3}{\beta};-\left(\frac{r}{r_{Sv}}\right)^{\beta}\right)+(\alpha+3) \delta_{c} \,  _2F_1\left(1,\frac{3}{\beta};\frac{\beta+3}{\beta};-\left(\frac{r}{r_{Sv}}\right)^{\beta}\right)\nonumber\\&+\alpha+3\Biggl)-\Lambda r_{0}^3 \rho_{a} \Biggl(-3 \delta_{c} \left(\frac{r_{0}}{r_{Sc}}\right)^{\alpha} \, _2F_1\left(1,\frac{\alpha+3}{\beta};\frac{\alpha+\beta+3}{\beta};-\left(\frac{r_{0}}{r_{Sv}}\right)^{\beta}\right)+(\alpha+3) \delta_{c} \, \nonumber\\& _2F_1\left(1,\frac{3}{\beta};\frac{\beta+3}{\beta};-\left(\frac{r_{0}}{r_{Sv}}\right)^{\beta}\right)+\alpha+3\Biggl)+3 (\alpha+3) r_{0}\Biggl],\label{p_r}
    \end{align}
    \begin{align}
    P_t(r)&=-\frac{1}{6 (\alpha+3) \Lambda r^3}\Biggl[3 \Lambda \delta_{c} r^3 \rho_{a} \left(\frac{r}{r_{Sc}}\right)^{\alpha} \, _2F_1\left(1,\frac{\alpha+3}{\beta};\frac{\alpha+\beta+3}{\beta};-\left(\frac{r}{r_{Sv}}\right)^{\beta}\right)-(\alpha+3) \Lambda \delta_{c} r^3 \rho_{a} \, _2F_1\left(1,\frac{3}{\beta};\frac{\beta+3}{\beta};-\left(\frac{r}{r_{Sv}}\right)^{\beta}\right)\nonumber\\&-3 \Lambda \delta_{c} r_{0}^3 \rho_{a} \left(\frac{r_{0}}{r_{Sc}}\right)^{\alpha} \, _2F_1\left(1,\frac{\alpha+3}{\beta};\frac{\alpha+\beta+3}{\beta};-\left(\frac{r_{0}}{r_{Sv}}\right)^{\beta}\right)+(\alpha+3) \Lambda \delta_{c} r_{0}^3 \rho_{a} \, _2F_1\left(1,\frac{3}{\beta};\frac{\beta+3}{\beta};-\left(\frac{r_{0}}{r_{Sv}}\right)^{\beta}\right)\nonumber\\&-\frac{3 \alpha \Lambda \delta_{c} r^3 \rho_{a} \left(\frac{r}{r_{Sc}}\right)^{\alpha}}{\left(\frac{r}{r_{Sv}}\right)^{\beta}+1}-\frac{9 \Lambda \delta_{c} r^3 \rho_{a} \left(\frac{r}{r_{Sc}}\right)^{\alpha}}{\left(\frac{r}{r_{Sv}}\right)^{\beta}+1}+\frac{3 \alpha \Lambda \delta_{c} r^3 \rho_{a}}{\left(\frac{r}{r_{Sv}}\right)^{\beta}+1}+2 \alpha \Lambda r^3 \rho_{a}+\alpha \Lambda r_{0}^3 \rho_{a}\nonumber\\&-3 \alpha r_{0}+\frac{9 \Lambda \delta_{c} r^3 \rho_{a}}{\left(\frac{r}{r_{Sv}}\right)^{\beta}+1}+6 \Lambda r^3 \rho_{a}+3 \Lambda r_{0}^3 \rho_{a}-9 r_{0}\Biggl].\label{p_t}
\end{align}
\end{small}
\end{widetext}

In GR, the weak energy condition is a key principle that provides insights into how energy behaves in the universe. When the weak energy condition is violated, it often indicates the presence of exotic matter--unusual forms of matter that don't follow our typical expectations. This violation serves as a warning sign, suggesting that the usual laws governing energy might not apply. Several energy conditions, including the null energy condition (NEC), weak energy condition WEC, dominant energy condition (DEC), and strong energy condition (SEC), act as important guidelines. They help ensure that our models of spacetime are consistent and realistic. By applying these mathematical rules to the energy-momentum tensor, we can assess whether certain spacetime scenarios, like wormholes, are physically possible:
\begin{eqnarray}
    T_{\xi \varrho}u^{\xi} u^{\varrho}\geq 0.
\end{eqnarray}
Here, we consider the energy-momentum tensor, denoted as $T_{\xi \varrho}$, and a timelike vector field represented by $u^\xi$. In their research on wormholes that use exotic matter, Hochberg and Visser \cite{Hochberg:1998ha, Hochberg:1998ii} built upon the important work done by Morris and Thorne \cite{Morris:1988cz}. They made a crucial discovery regarding the NEC, showing that it can actually be violated. Specifically, they found that the throat region of a wormhole does not adhere to the NEC. To gain a deeper understanding of the dynamics involved, we can look at the Raychaudhuri equations. These equations help us understand how certain properties change over time, including:
\begin{itemize}
    \item \textit{Expansion}: $ \Omega $,
    \item \textit{Rotation}: $ w_{\xi \varrho} $,
    \item \textit{Shear}: $ \sigma_{\xi \varrho} $.
\end{itemize}
These properties relate to the paths defined by the timelike vector $ v^\xi $, and the Raychaudhuri equations can be expressed in the following way:
\begin{eqnarray}
\frac{d \Omega}{d \tau} &=& - \frac{\Omega^2}{3}-R_{\xi \varrho} v^{\xi} v^{\varrho}-\sigma_{\xi \varrho} \sigma^{\xi \varrho}-w_{\xi \varrho} w^{\xi \varrho},\label{aa}\\
\frac{d \Omega}{d \tau} &=&- \frac{\Omega^2}{2}-R_{\xi \varrho} u^{\xi} u^{\varrho}-\sigma_{\xi \varrho} \sigma^{\xi \varrho}-w_{\xi \varrho} w^{\xi \varrho}.
\end{eqnarray}
Exploring modified gravity opens up fascinating insights into energy conditions, especially through the Raychaudhuri equation \cite{IV: RAY}. This equation relates to an effective stress-energy tensor and helps us understand how changes in gravitational theory can influence the behavior of matter and energy in the universe. If we're looking for a more in-depth discussion on these energy conditions, particularly regarding $f(R)$ gravity, we highly recommend checking out \cite{IV: CAP}. When it comes to wormholes, the NEC plays a vital role. Essentially, this condition tells us that the combined energy density and pressure should be non-negative. However, if we find that the NEC is violated at the throat of a wormhole, it suggests the presence of exotic matter--this is the kind of matter that can have negative energy density \cite{Morris:1988cz}. To clarify this further, we can look at the expressions for energy density, denoted as $\rho$, along with radial pressure $P_r$ and tangential pressure $P_t$, as outlined in Eq. \eqref{14a}-\eqref{14c}. By examining these quantities at a specific point, $r = r_0$, we can reformulate the NEC and delve into what this means for the stability and possibility of wormholes in scenarios involving modified gravity.
\begin{widetext}
\begin{align}
\begin{split}
\label{NECr}
 \rho+P_r\bigg\vert_{r=r_0}& =  \frac{\rho_{a} \left(-\delta_{c} \left(\frac{r_{0}}{r_{Sc}}\right)^{\alpha}+\left(\frac{r_{0}}{r_{Sv}}\right)^{\beta}+\delta_{c}+1\right)}{\left(\frac{r_{0}}{r_{Sv}}\right)^{\beta}+1}-\frac{1}{\Lambda r_{0}^2} \overset{!}{\geq}0\,\\
 \rho+P_t\bigg\vert_{r=r_0} &=  \frac{\Lambda r_{0}^2 \rho_{a} \left(-\delta_{c} \left(\frac{r_{0}}{r_{Sc}}\right)^{\alpha}+\left(\frac{r_{0}}{r_{Sv}}\right)^{\beta}+\delta_{c}+1\right)+\left(\frac{r_{0}}{r_{Sv}}\right)^{\beta}+1}{2 \Lambda r_{0}^2 \left(\left(\frac{r_{0}}{r_{Sv}}\right)^{\beta}+1\right)}\overset{!}{\geq}0\,.
        \end{split}
\end{align}
\end{widetext}
Let's shift our focus to the other energy conditions for a moment. A crucial point to remember is that whenever the NEC is applicable, the WEC also needs to be met, especially since we're considering $\rho$ to be positive. If we take a closer look at Eqs. \eqref{15a}-\eqref{15c}, we can obtain
\begin{equation}
\label{eq:SEC}
    \rho+P_r+2P_t=0\,.
\end{equation}
This tells us that the SEC is completely dependent on the NEC. To see if we can draw a similar conclusion for the DEC, we can express
\begin{widetext}
\begin{align}
\begin{split}
    \rho-|P_r|\bigg\vert_{r=r_0}&=\frac{\rho_{a} \left(-\delta_{c} \left(\frac{r_{0}}{r_{Sc}}\right)^{\alpha}+\left(\frac{r_{0}}{r_{Sv}}\right)^{\beta}+\delta_{c}+1\right)}{\left(\frac{r_{0}}{r_{Sv}}\right)^{\beta}+1}+\frac{1}{\Lambda r_{0}^2}\\
     \rho-|P_t|\bigg\vert_{r=r_0}&=-\frac{-3 \Lambda r_{0}^2 \rho_{a} \left(-\delta_{c} \left(\frac{r_{0}}{r_{Sc}}\right)^{\alpha}+\left(\frac{r_{0}}{r_{Sv}}\right)^{\beta}+\delta_{c}+1\right)+\left(\frac{r_{0}}{r_{Sv}}\right)^{\beta}+1}{2 \Lambda r_{0}^2 \left(\left(\frac{r_{0}}{r_{Sv}}\right)^{\beta}+1\right)}.
\end{split}
\end{align}
\end{widetext}

\begin{figure*}
\centering
\includegraphics[width=8.3cm,height=5.75cm]{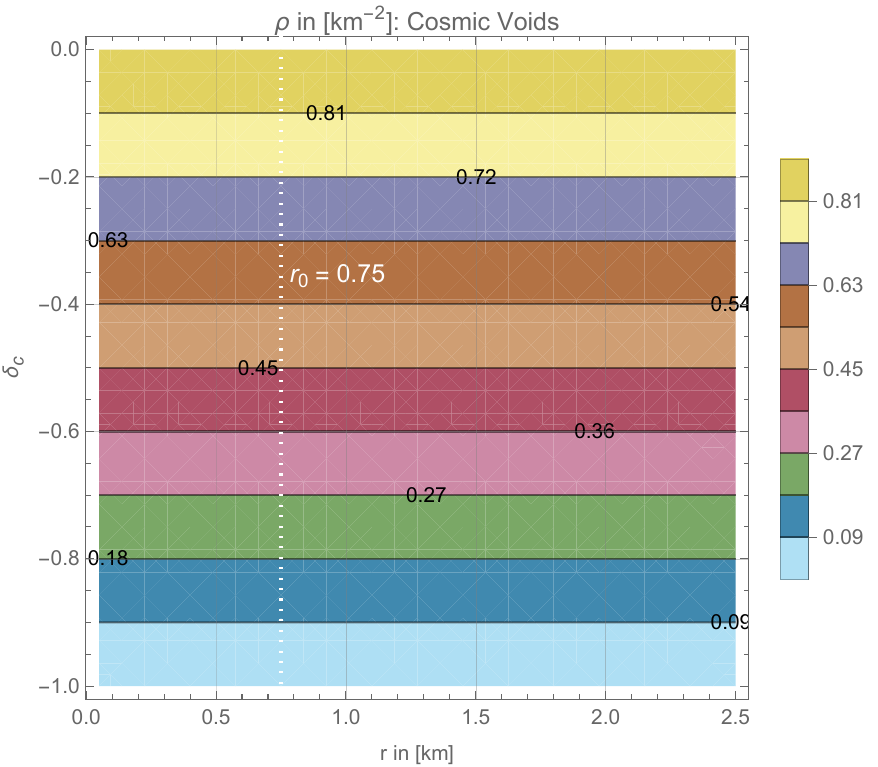}
\caption{For cosmic voids wormholes, the matter density, $\rho$, is characterized by a constant throat radius of $r_0 = 0.75$, applicable for various negative values of the parameter $\delta_c$. Furthermore, several key parameters are crucial: $\lambda = 0.75$, $\chi = 0.75$, $\alpha = 3.75$, $\beta = 6.5$, $\rho_{a} = 0.00001$, $r_{Sc} = 75$, and $r_{Sv} = 95$. }\label{fig2}
\end{figure*}

\begin{figure*}
\centering
\includegraphics[width=8.3cm,height=5.75cm]{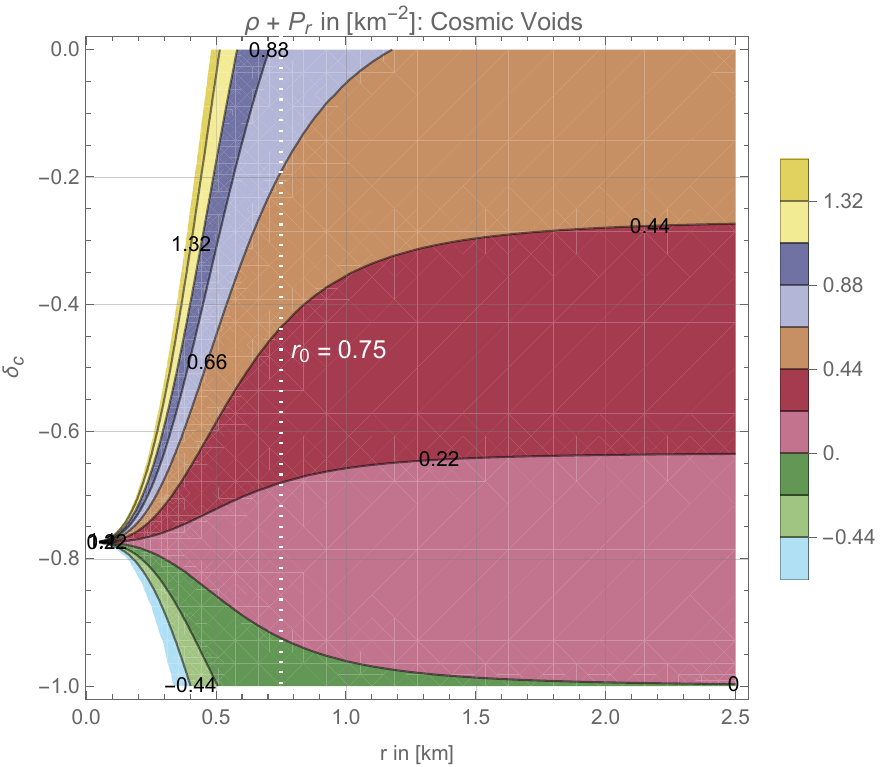}
\includegraphics[width=8.3cm,height=5.75cm]{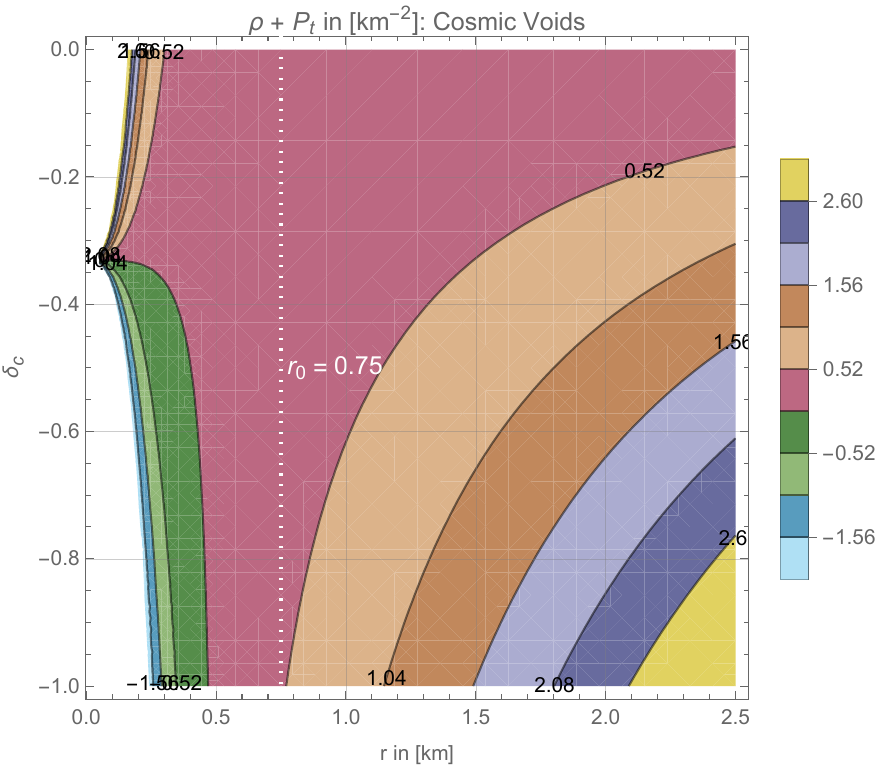}
\caption{For cosmic voids wormholes, the  NEC, $\rho + P_r$ and $\rho + P_t$, is characterized by a constant throat radius of $r_0 = 0.75$, applicable for various negative values of the parameter $\delta_c$. Furthermore, several key parameters are crucial: $\lambda = 0.75$, $\chi = 0.75$, $\alpha = 3.75$, $\beta = 6.5$, $\rho_{a} = 0.00001$, $r_{Sc} = 75$, and $r_{Sv} = 95$. }\label{fig3}
\end{figure*}

\begin{figure*}[ht]
\centering
\includegraphics[width=8.3cm,height=5.75cm]{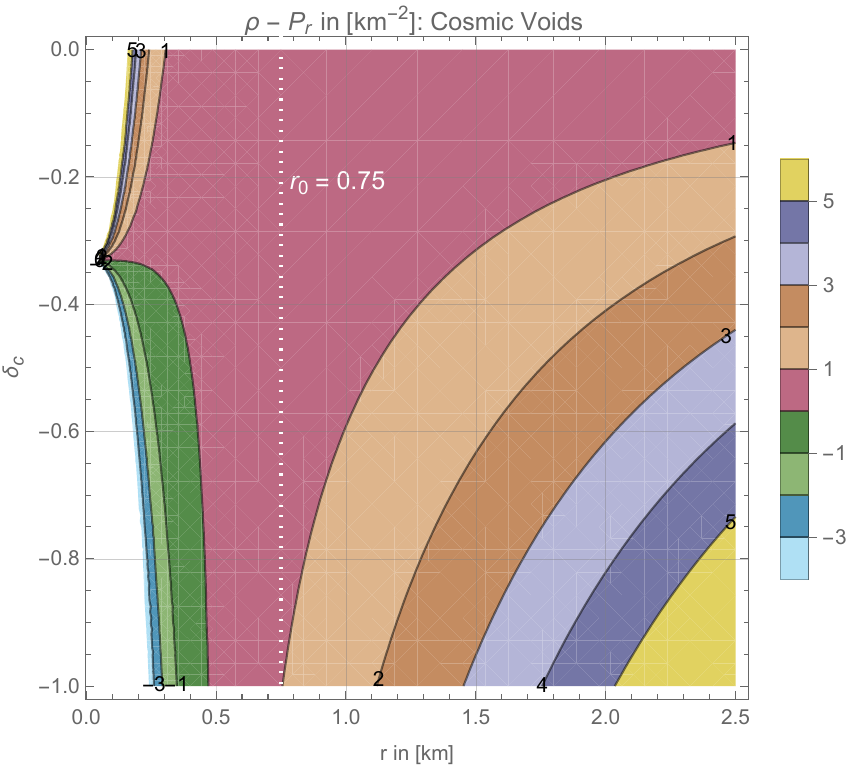}
\includegraphics[width=8.3cm,height=5.75cm]{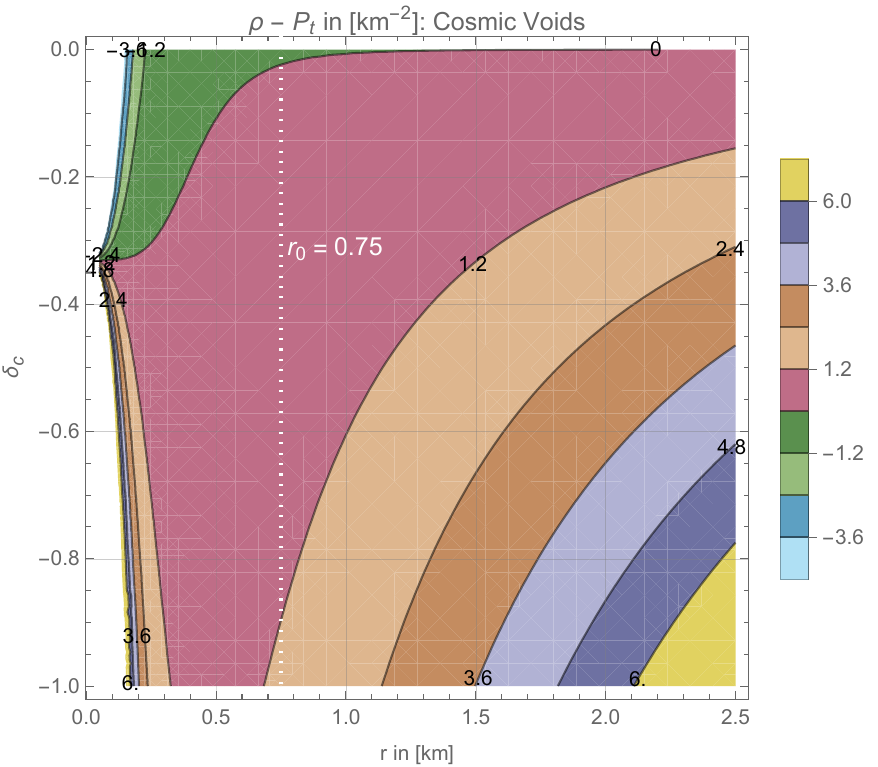}
\caption{ For cosmic voids wormholes, the  DEC, $\rho - P_r$ and $\rho - P_t$, is characterized by a constant throat radius of $r_0 = 0.75$, applicable for various negative values of the parameter $\delta_c$. Furthermore, several key parameters are crucial: $\lambda = 0.75$, $\chi = 0.75$, $\alpha = 3.75$, $\beta = 6.5$, $\rho_{a} = 0.00001$, $r_{Sc} = 75$, and $r_{Sv} = 95$. }\label{fig5}
\end{figure*}

During our analysis, we focused on the parameter $\delta_{c}$, which is associated with cosmic voids. We observe that the energy density remains positive throughout the spacetime for all considered values of $\delta_{c} \in [-1, 0]$, based on the parameter set $r_0 = 0.75$, $\lambda = 0.75$, $\chi = 0.75$, $\alpha = 3.75$, $\beta = 6.5$, $\rho_a = 0.00001$, $r_{Sc} = 75$, and $r_{Sv} = 95$, as shown in Fig.~\ref{fig2}. This persistent positivity ensures that the model avoids unphysical behavior and maintains physical plausibility. The parameter $\delta_{c}$ plays a critical role in tuning the gravitational interactions, which in turn influence the energy conditions. Even for negative values of $\delta_{c}$, the energy density stays positive, indicating that the system adheres to standard energy requirements. Specifically, the radial NEC, $\rho + P_r$, is satisfied for $\delta_{c} \in (-0.925, 0]$, but violated for $\delta_{c} \in [-1, -0.925)$, as illustrated in Fig.~\ref{fig3} (left panel). This implies that for a significant portion of the parameter space, the radial energy flow remains physically acceptable. Simultaneously, the tangential NEC, $\rho + P_t$, is satisfied at the throat and continues to hold beyond it for all considered $\delta_{c}$, as shown in Fig.~\ref{fig3} (right panel). The full satisfaction of the tangential NEC suggests that energy flow in the angular directions is positive, reinforcing the potential absence of exotic matter in those directions. The radial DEC $\rho - P_r$ is fully satisfied both at the throat and in the exterior region (see Fig.~\ref{fig5} (left panel)), supporting the dominance of energy density over radial pressure. In contrast, the tangential DEC $\rho - P_t$ (see Fig.~\ref{fig5} (right panel)) is violated in the range $\delta_{c} \in (-0.027, 0]$, though it holds for $\delta_{c} \in [-1, -0.027]$, showing some sensitivity to the parameter variation in the tangential direction. Finally, the SEC, expressed as $\rho + P_r + 2P_t = 0$, is satisfied in this setup. This indicates that the model remains consistent with GR's prediction for how matter and energy influence spacetime geometry. Although exotic matter cannot be entirely ruled out, introducing $\delta_{c}$ as a free parameter tied to cosmic voids helps to minimize or localize energy condition violations. This flexibility opens the door to viable traversable wormhole configurations within a physically reasonable framework.


\section{Gravitational Lensing Effects}
\label{ch: V}
In this section, we'll explore the fascinating world of wormholes, including different types like charged, massless, Janis-Newman-Winnicour, and Ellis wormholes \cite{V: GOD, V: DEY, V: TSU}. We'll particularly focus on their lensing effects, which provide important insights into the structure of black holes and naked singularities \cite{V: VIR, V: VIR1, V: BOZZA, V: BOZZA1}. We're adopting a well-established approach from the literature to investigate how light-like particles behave as they approach the throat of a wormhole--an intriguing phenomenon known as \textit{gravitational lensing}. If we're interested in the detailed derivations of the formulas we'll be discussing, we can find more information in \cite{V: WEI}. To get started, let's define the Lagrangian associated with a null curve $\gamma(s) = \bigl(t(s), r(s), \theta(s), \phi(s)\bigr)$, which exists within the spacetime described by Eq. (\ref{1}), where $\Dot{\gamma} = \frac{\partial \gamma}{\partial s}$:
\begin{align}
    \mathcal{L}=-e^{-2\Phi}\frac{\Dot{t}^2}{2}+\frac{1}{1-b/r}\frac{\Dot{r}^2}{2}+r^2\frac{\Dot{\theta}^2}{2}+r^2\sin^2{\theta}\frac{\Dot{\phi}^2}{2}=0.
\end{align}
Let's take a closer look at the effective potential $V_{eff}(r)$, which we can define as $V_{eff}(r) := \frac{e^{2\Phi} L^2}{r^2}$. This expression emerges from our analysis of the conserved quantities $E$ and $L$. Specifically, we define $E$ as $ E := \frac{\partial \mathcal{L}}{\partial \Dot{t}} = e^{-2\Phi} \Dot{t}$, and $L$ as $ L := \frac{\partial \mathcal{L}}{\partial \Dot{\phi}} = r^2 \Dot{\phi}$. With these definitions, we can express our equation in a more insightful way:
$ E^2 = \frac{e^{2\Phi}}{1 - \hat{X}/r} \Dot{r}^2 + V_{eff}(r)$. The symmetries in the coordinates $t$ and $\phi$ really enhance our understanding of the dynamics involved. The spherical symmetry of the line element $ds^2$ shows that $\Dot{\theta} = 0$, which simplifies our analysis by allowing us to set $\theta = \frac{\pi}{2}$.

Let's focus on the circular orbits of light-like paths in the effective potential $V_{\text{eff}}$, often referred to as \textit{photon spheres} \cite{V: VIR2}. This also brings energy $E$ and angular momentum $L$ into consideration. When we examine the minima of $V_{\text{eff}}$, we find that $\frac{dV_{\text{eff}}}{dr} \neq 0$ at a constant redshift, which can be a bit puzzling. This confusion usually arises from a less-than-ideal choice of coordinate system. By switching to proper coordinates $l(r)$ at $r_0$, we can see that two different values of $l$ can correspond to the same $r$. This highlights the connection between the regions of spacetime linked by the wormhole, making everything much clearer, $l(r)=\pm\bigintss_{\,r_0}^r\frac{1}{\sqrt{1-\frac{\hat{X}'}{r'}}}\text{d}r'$. We find that the potential peaks at $r = r_0$ (where $l = 0$) \cite{V: SHA}. This means the throat of the wormhole behaves like an unstable photon sphere. It's an interesting point that we can explore further by examining the angle $\alpha(r_{tp})$, which shows how much the photon gets deflected:
\begin{align}
\alpha(r_{tp}) = -\pi + 2 \int_{r_{tp}}^\infty \frac{e^\Phi}{\sqrt{1 - \frac{\hat{X}'}{r'}} \sqrt{\frac{r^2}{\mathcal{K}^2} - e^{2\Phi}}} \text{d}r.
\label{eq: DA}
\end{align}
Here, $\mathcal{K} := L/E$ is known as the \textit{impact parameter}. When we look at a constant redshift, we see that $\mathcal{K}$ is directly related to the turning point $r_{tp}$, where the rate of change $\dot{r}(r_{tp})$ equals zero: $\mathcal{K} = r_{tp} e^{-\Phi}$. In simpler terms, $\alpha(r_{tp})$ depends on the ratio $L/E$. Since $r = r_0$ is the only photon sphere in this scenario, the only singularity in Eq. (\ref{eq: DA}) occurs at $r_{tp} = r_0$. For values of $r_{tp} < r_0$, the function diverges. A growing body of research has explored how light behaves near wormholes and what this could mean for real-world observations. Unlike black holes, which bend light inward due to their intense gravitational pull, wormholes can cause light to bend outward near their throats---a strikingly different and repulsive effect \cite{Nandi:2008ij, Dey:2008kn}. This reversal in behavior might manifest in the sky as unusual lensing effects: fewer or oddly positioned images, unexpected asymmetries, or even missing arcs where standard theory predicts them \cite{Tsukamoto:2016zdu}. Additionally, wormholes can cast unique shadow patterns that don't resemble the classic black hole silhouette, offering potential clues for future imaging efforts using very-long-baseline interferometry \cite{Shaikh:2018kfv}. Light traveling near a wormhole doesn't always follow the rules we expect---it can arrive earlier or later than usual, thanks to the warped geometry around the throat \cite{Perlick:2004tq}. In some cases, if a wormhole acts as a lens, it could produce strange microlensing signals---dimmer, lopsided, or unusually shaped light curves that stand out from the smooth, symmetric ones produced by ordinary stars or black holes \cite{Abe:2010ap, Toki:2011zu, Safonova:2001vz}. Although most plasma studies focus on black hole environments, similar frameworks can be extended to wormholes and may subtly alter their observational signatures \cite{Perlick:2015vta}. Altogether, these findings suggest that wormholes, if they exist, could leave behind clear, detectable imprints in the light they deflect.

Now, let's take a closer look at the numerical results for various values of the parameter $\delta{c}$. Now, we can explore the numerical results for different values of the parameter $\delta_{c}$. In Fig. \ref{fig9}, we see how the deflection angle $\alpha(r_{tp})$ relates to the parameters $\delta_{c}$ and $r_{tp}$. When $\delta_{c}$ is between $-1$ and $0$, the deflection angle remains negative for all corresponding values of $\alpha(r_{tp})$. This indicates that photons experience a repulsive gravitational force, pushing them away from the throat of the wormhole rather than drawing them in. The figure clearly illustrates how the parameter $\delta_{c}$, influenced by cosmic voids, plays a significant role in shaping the deflection angle and the overall characteristics of the wormhole.

\begin{figure*}[ht]
\centering
\includegraphics[width=8.3cm,height=5.75cm]{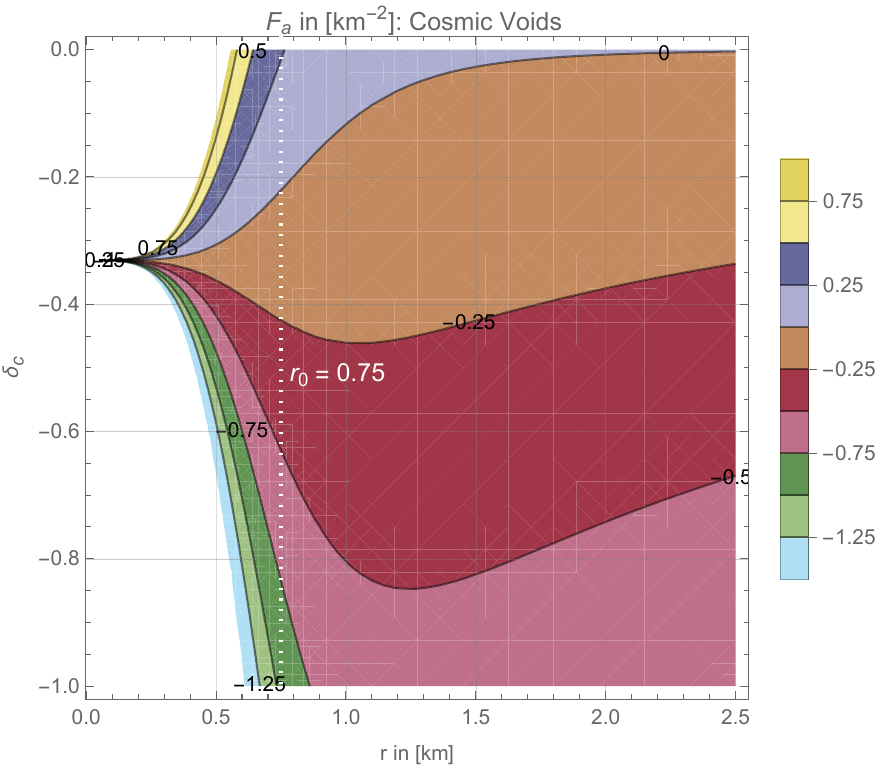}
\includegraphics[width=8.3cm,height=5.75cm]{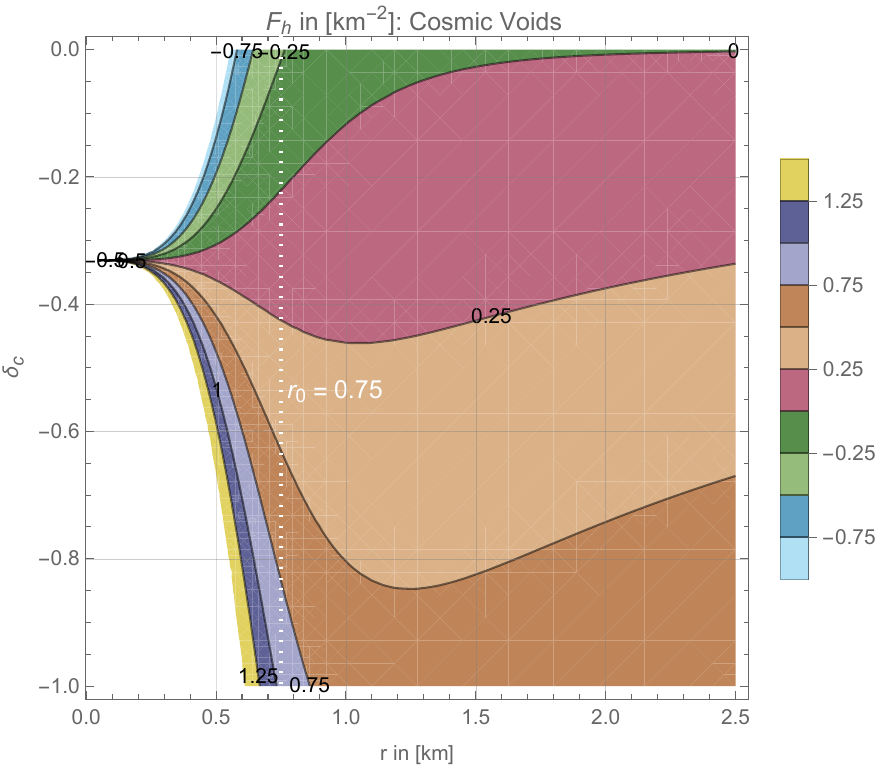}
\caption{ For cosmic voids wormholes, the  anisotropic and hydrostatic forces, $F_a$ and $F_h$, are characterized by a constant throat radius of $r_0 = 0.75$, applicable for various negative values of the parameter $\delta_c$. Furthermore, several key parameters are crucial: $\lambda = 0.75$, $\chi = 0.75$, $\alpha = 3.75$, $\beta = 6.5$, $\rho_{a} = 0.00001$, $r_{Sc} = 75$, and $r_{Sv} = 95$. }\label{fig6}
\end{figure*}
\begin{figure*}[ht]
\centering
\includegraphics[width=8.3cm,height=5.75cm]{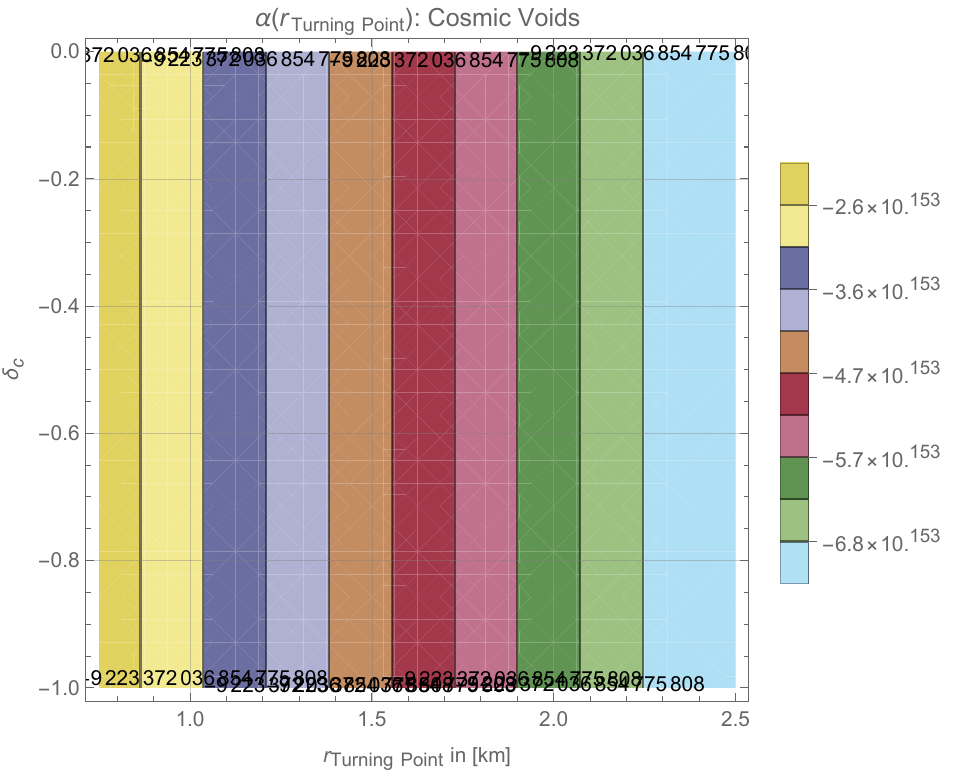}
\caption{ For cosmic voids wormholes, the  deflection angle, $\alpha(r_{tp})$, are characterized by a constant throat radius of $r_0 = 0.75$, applicable for various negative values of the parameter $\delta_c$. Furthermore, several key parameters are crucial: $\lambda = 0.75$, $\chi = 0.75$, $\alpha = 3.75$, $\beta = 6.5$, $\rho_{a} = 0.00001$, $r_{Sc} = 75$, and $r_{Sv} = 95$. }\label{fig9}
\end{figure*}

\begin{figure*}[ht]
\centering
\includegraphics[width=8.3cm,height=5.75cm]{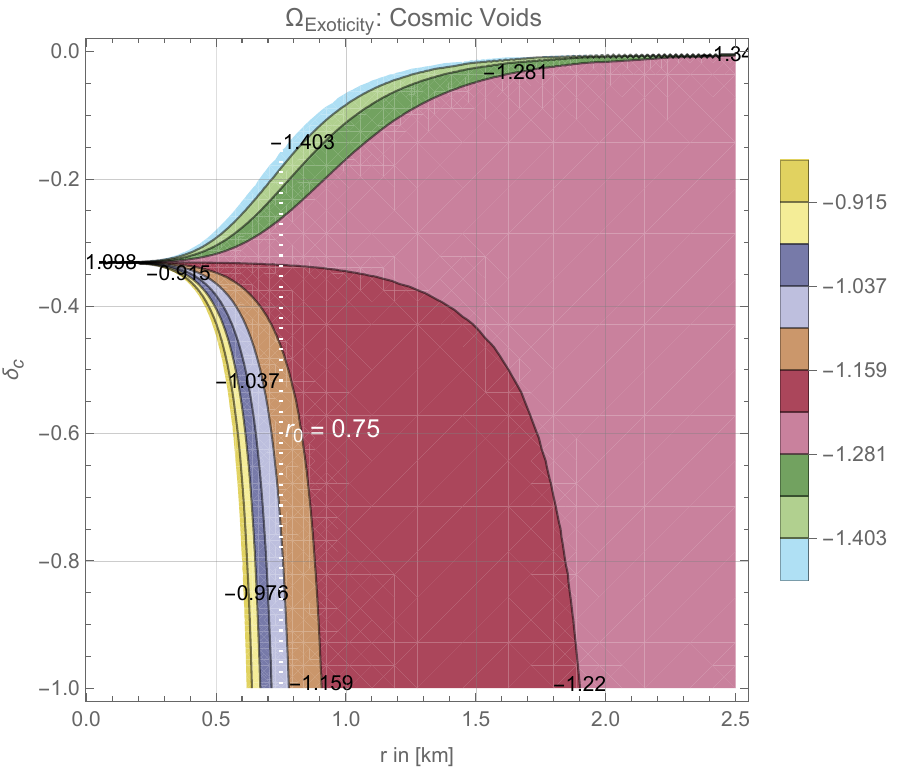}
\includegraphics[width=8.3cm,height=5.75cm]{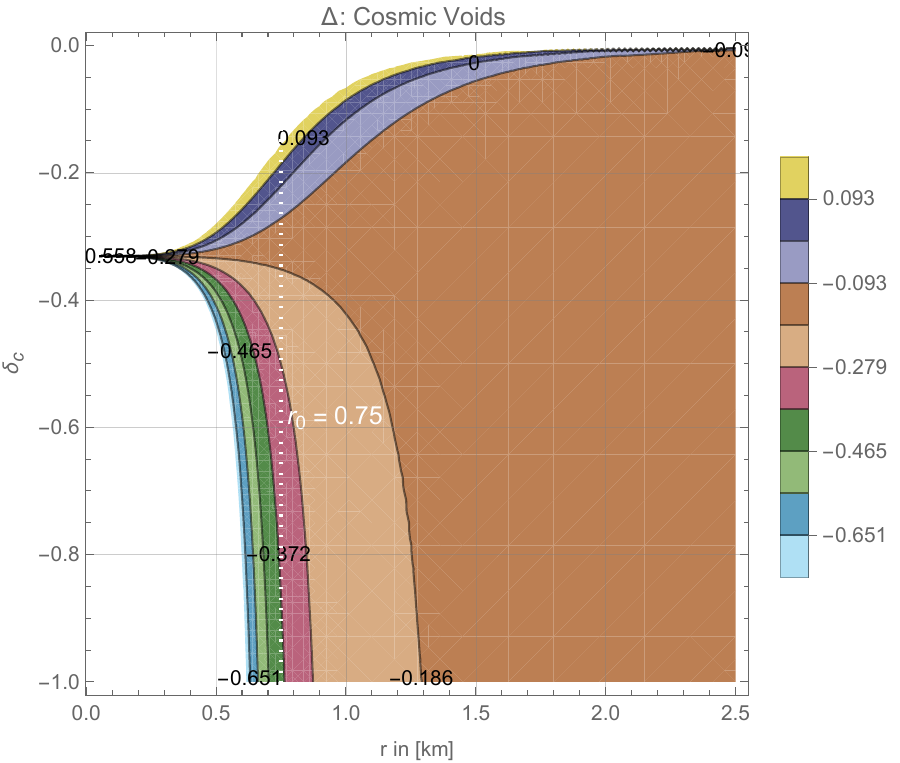}
\caption{ For cosmic voids wormholes, the exoticity factor and anisotropy parameter, $\Omega_{Exoticity}$ and $\Delta$, is characterized by a constant throat radius of $r_0 = 0.75$, applicable for various negative values of the parameter $\delta_c$. Furthermore, several key parameters are crucial: $\lambda = 0.75$, $\chi = 0.75$, $\alpha = 3.75$, $\beta = 6.5$, $\rho_{a} = 0.00001$, $r_{Sc} = 75$, and $r_{Sv} = 95$. }\label{fig7}
\end{figure*} 

\section{Tolman-Oppenheimer-Volkoff (TOV) Criteria}\label{ch: VI}
The TOV equation is an essential tool for understanding how gravitational equilibrium works in spherically symmetric spacetimes. It helps us evaluate the stability of various solutions, not just in GR but also in modified gravity scenarios \cite{Gorini}. Additionally, this equation can be adapted to account for anisotropic mass distributions in standard GR \cite{Kuhfittig}. By generalizing the TOV equation, we gain a more comprehensive framework for exploring the behavior of these fascinating systems under different conditions
\begin{eqnarray}\label{51}
\frac{\nu'}{2}(\rho+P_r)+\frac{dP_r}{dr}+\frac{2}{r}(P_r-P_t)=0.
\end{eqnarray}
In our exploration, we've noticed in Eq. \eqref{1} that $\tilde{\nu}(r) = 2\Phi(r)$. This insight helps us better grasp the various forces involved, such as anisotropic force, gravitational force, and hydrostatic pressure
\begin{equation}\label{52}
F_h=-\frac{dP_r}{dr}, ~~F_g=-\frac{\nu'}{2}(\rho+P_r), ~F_a=\frac{2}{r}(P_t-P_r),    
\end{equation}
By applying a constant redshift and recognizing that we reach equilibrium when the equation $F_h + F_g + F_a = 0$ holds true, we can simplify our analysis and exclude $F_g$ from our calculations
\begin{equation}
\label{eq: eq}
    F_h+F_a=0.
\end{equation}
Within our framework of linearly modified gravity, the dynamic characteristics of the energy-momentum tensor introduce additional terms \cite{I: HAG} 
\begin{equation}
    \nabla^\mu T_{\mu\nu}=-\frac{1}{\kappa}\Biggl[\Bigl(\frac{\eta}{2}+\chi\Bigr)\nabla_\nu\rho+\frac{1}{2}\Bigl(\chi\nabla_\nu T-\frac{\eta}{2}\nabla_\nu \rho\Bigr)\Biggr].
\end{equation}
The force $F_m$ can be represented as \cite{VI: DGU}  
\begin{equation}
    F_m=\frac{\chi}{2\kappa}(\rho'+P'_r+2P'_t).
\end{equation}
Nonetheless, this force is ultimately zero everywhere, given that $\rho + P_r + 2P_t = 0$, as shown in Eq. \eqref{eq:SEC}. Therefore, we will use Eq. \eqref{eq: eq} as the fundamental condition for the stability of the solutions under discussion. We will begin by examining the anisotropic force as defined in Eq. \eqref{52}. Next, we will analyze Eqs. \eqref{p_r} and \eqref{p_t} and make the necessary adjustments. Finally, we will compute the hydrostatic force by differentiating Eq. \eqref{p_r}, which will lead us to:
\begin{widetext}
\begin{align}
F_a&=\frac{1}{(\alpha+3) \Lambda r^4 \left(\left(\frac{r}{r_{Sv}}\right)^{\beta}+1\right)}\Biggl[-3 \Lambda \delta_{c} r^3 \rho_{a} \left(\frac{r}{r_{Sc}}\right)^{\alpha} \left(\frac{r}{r_{Sv}}\right)^{\beta} \, _2F_1\left(1,\frac{\alpha+3}{\beta};\frac{\alpha+\beta+3}{\beta};-\left(\frac{r}{r_{Sv}}\right)^{\beta}\right)-3 \Lambda \delta_{c} r^3 \rho_{a} \left(\frac{r}{r_{Sc}}\right)^{\alpha} \,\nonumber\\& _2F_1\left(1,\frac{\alpha+3}{\beta};\frac{\alpha+\beta+3}{\beta};-\left(\frac{r}{r_{Sv}}\right)^{\beta}\right)+(\alpha+3) \Lambda \delta_{c} r^3 \rho_{a} \left(\left(\frac{r}{r_{Sv}}\right)^{\beta}+1\right) \, _2F_1\left(1,\frac{3}{\beta};\frac{\beta+3}{\beta};-\left(\frac{r}{r_{Sv}}\right)^{\beta}\right)\nonumber\\&+3 \Lambda \delta_{c} r_{0}^3 \rho_{a} \left(\frac{r_{0}}{r_{Sc}}\right)^{\alpha} \left(\frac{r}{r_{Sv}}\right)^{\beta} \, _2F_1\left(1,\frac{\alpha+3}{\beta};\frac{\alpha+\beta+3}{\beta};-\left(\frac{r_{0}}{r_{Sv}}\right)^{\beta}\right)-(\alpha+3) \Lambda \delta_{c} r_{0}^3 \rho_{a} \left(\left(\frac{r}{r_{Sv}}\right)^{\beta}+1\right) \,\nonumber\\& _2F_1\left(1,\frac{3}{\beta};\frac{\beta+3}{\beta};-\left(\frac{r_{0}}{r_{Sv}}\right)^{\beta}\right)+3 \Lambda \delta_{c} r_{0}^3 \rho_{a} \left(\frac{r_{0}}{r_{Sc}}\right)^{\alpha} \, _2F_1\left(1,\frac{\alpha+3}{\beta};\frac{\alpha+\beta+3}{\beta};-\left(\frac{r_{0}}{r_{Sv}}\right)^{\beta}\right)\nonumber\\&-\alpha \Lambda r_{0}^3 \rho_{a} \left(\frac{r}{r_{Sv}}\right)^{\beta}+3 \alpha r_{0} \left(\frac{r}{r_{Sv}}\right)^{\beta}+\alpha \Lambda \delta_{c} r^3 \rho_{a} \left(\frac{r}{r_{Sc}}\right)^{\alpha}+3 \Lambda \delta_{c} r^3 \rho_{a} \left(\frac{r}{r_{Sc}}\right)^{\alpha}-\alpha \Lambda \delta_{c} r^3 \rho_{a}\nonumber\\&-\alpha \Lambda r_{0}^3 \rho_{a}+3 \alpha r_{0}-3 \Lambda r_{0}^3 \rho_{a} \left(\frac{r}{r_{Sv}}\right)^{\beta}+9 r_{0} \left(\frac{r}{r_{Sv}}\right)^{\beta}-3 \Lambda \delta_{c} r^3 \rho_{a}-3 \Lambda r_{0}^3 \rho_{a}+9 r_{0}\Biggl],
\end{align}
\begin{align}
        F_h &= \frac{1}{(\alpha+3) \Lambda r^4 \left(\left(\frac{r}{r_{Sv}}\right)^{\beta}+1\right)}\Biggl[3 \Lambda \delta_{c} r^3 \rho_{a} \left(\frac{r}{r_{Sc}}\right)^{\alpha} \left(\frac{r}{r_{Sv}}\right)^{\beta} \, _2F_1\left(1,\frac{\alpha+3}{\beta};\frac{\alpha+\beta+3}{\beta};-\left(\frac{r}{r_{Sv}}\right)^{\beta}\right)+3 \Lambda \delta_{c} r^3 \rho_{a} \left(\frac{r}{r_{Sc}}\right)^{\alpha} \, \nonumber\\& _2F_1\left(1,\frac{\alpha+3}{\beta};\frac{\alpha+\beta+3}{\beta};-\left(\frac{r}{r_{Sv}}\right)^{\beta}\right)-(\alpha+3) \Lambda \delta_{c} r^3 \rho_{a} \left(\left(\frac{r}{r_{Sv}}\right)^{\beta}+1\right) \, _2F_1\left(1,\frac{3}{\beta};\frac{\beta+3}{\beta};-\left(\frac{r}{r_{Sv}}\right)^{\beta}\right)\nonumber\\&-3 \Lambda \delta_{c} r_{0}^3 \rho_{a} \left(\frac{r_{0}}{r_{Sc}}\right)^{\alpha} \left(\frac{r}{r_{Sv}}\right)^{\beta} \, _2F_1\left(1,\frac{\alpha+3}{\beta};\frac{\alpha+\beta+3}{\beta};-\left(\frac{r_{0}}{r_{Sv}}\right)^{\beta}\right)+(\alpha+3) \Lambda \delta_{c} r_{0}^3 \rho_{a} \left(\left(\frac{r}{r_{Sv}}\right)^{\beta}+1\right) \nonumber\\& \, _2F_1\left(1,\frac{3}{\beta};\frac{\beta+3}{\beta};-\left(\frac{r_{0}}{r_{Sv}}\right)^{\beta}\right)-3 \Lambda \delta_{c} r_{0}^3 \rho_{a} \left(\frac{r_{0}}{r_{Sc}}\right)^{\alpha} \, _2F_1\left(1,\frac{\alpha+3}{\beta};\frac{\alpha+\beta+3}{\beta};-\left(\frac{r_{0}}{r_{Sv}}\right)^{\beta}\right)\nonumber\\&+\alpha \Lambda r_{0}^3 \rho_{a} \left(\frac{r}{r_{Sv}}\right)^{\beta}-3 \alpha r_{0} \left(\frac{r}{r_{Sv}}\right)^{\beta}-\alpha \Lambda \delta_{c} r^3 \rho_{a} \left(\frac{r}{r_{Sc}}\right)^{\alpha}-3 \Lambda \delta_{c} r^3 \rho_{a} \left(\frac{r}{r_{Sc}}\right)^{\alpha}+\alpha \Lambda \delta_{c} r^3 \rho_{a}+\alpha \Lambda r_{0}^3 \rho_{a}\nonumber\\&-3 \alpha r_{0}+3 \Lambda r_{0}^3 \rho_{a} \left(\frac{r}{r_{Sv}}\right)^{\beta}-9 r_{0} \left(\frac{r}{r_{Sv}}\right)^{\beta}+3 \Lambda \delta_{c} r^3 \rho_{a}+3 \Lambda r_{0}^3 \rho_{a}-9 r_{0}\Biggl].
\end{align}
\end{widetext}
Upon closer examination of the involved forces, we find that the stability of the wormhole is supported by the TOV condition, expressed as $F_a + F_h = 0$. This indicates that the anisotropic force $F_a$ and the hydrostatic force $F_h$ are in perfect equilibrium. In Fig. \ref{fig6}, we can observe the interaction of these forces for cosmic voids wormholes. Notably, both forces are equal in magnitude but act in opposing directions, which aligns with our expectations for a stable configuration. This balance is crucial for maintaining the integrity of the wormhole structure.

\section{Exotic matter, exoticity parameter, and anisotropy parameter's influence on wormhole geometry} \label{ch: VII}
\subsection{Exotic matter and the exoticity parameter in wormholes}
To explore how exotic materials act near the neck of a wormhole, researchers utilize a concept known as the exoticity factor, as outlined by Lemos et al. in~\cite{Lemos:2003jb}. For a wormhole to allow travel through it, it needs to be filled with exotic matter, which is quite different from the regular matter we find in the universe. This exotic matter defies the rules of the NEC, meaning it doesn't meet the weak energy condition or other related criteria
\begin{widetext}
\begin{small}
\begin{align}
  \Omega_{Exoticity}=-\frac{\rho-P_r}{|\rho|}
&=-\frac{1}{\rho_{a} \left(-\delta_{c} \left(\frac{r}{r_{Sc}}\right)^{\alpha}+\left(\frac{r}{r_{Sv}}\right)^{\beta}+\delta_{c}+1\right)}\Biggl[\left(\left(\frac{r}{r_{Sv}}\right)^{\beta}+1\right) \Biggl(\frac{1}{3 (\alpha+3) \Lambda r^3}\Biggl[\Lambda r^3 \rho_{a} \Biggl(-3 \delta_{c} \left(\frac{r}{r_{Sc}}\right)^{\alpha} \, \nonumber\\& _2F_1\left(1,\frac{\alpha+3}{\beta};\frac{\alpha+\beta+3}{\beta};-\left(\frac{r}{r_{Sv}}\right)^{\beta}\right)+(\alpha+3) \delta_{c} \, _2F_1\left(1,\frac{3}{\beta};\frac{\beta+3}{\beta};-\left(\frac{r}{r_{Sv}}\right)^{\beta}\right)+\alpha+3\Biggl)\nonumber\\&-\Lambda r_{0}^3 \rho_{a} \Biggl(-3 \delta_{c} \left(\frac{r_{0}}{r_{Sc}}\right)^{\alpha} \, _2F_1\left(1,\frac{\alpha+3}{\beta};\frac{\alpha+\beta+3}{\beta};-\left(\frac{r_{0}}{r_{Sv}}\right)^{\beta}\right)+(\alpha+3) \delta_{c} \,\nonumber\\& _2F_1\left(1,\frac{3}{\beta};\frac{\beta+3}{\beta};-\left(\frac{r_{0}}{r_{Sv}}\right)^{\beta}\right)+\alpha+3\Biggl)+3 (\alpha+3) r_{0}\Biggl]+\frac{\rho_{a} \left(-\delta_{c} \left(\frac{r}{r_{Sc}}\right)^{\alpha}+\left(\frac{r}{r_{Sv}}\right)^{\beta}+\delta_{c}+1\right)}{\left(\frac{r}{r_{Sv}}\right)^{\beta}+1}\Biggl)\Biggl]
\end{align}
\end{small}
\end{widetext}
The negativity of the exoticity factor, $\Omega_{Exoticity}$, provides strong evidence that exotic matter gradually transforms into ordinary matter. This change highlights the dynamic nature of the environment surrounding the wormhole, illustrating how the properties of matter evolve with distance, as noted by Lemos et al.~\cite{Lemos:2003jb} and Kim et al.~\cite{Kim:2003zb}. When the exoticity factor becomes positive, it provides strong evidence that exotic matter is present near the throat of the wormhole, which is crucial for enabling travel through it. The left panel of Fig.~\ref{fig7} shows the exoticity parameter for cosmic voids in wormholes, revealing a negative trend. This suggests that there is no exotic matter present around the throats of these wormholes in the context of $f(R,\mathcal{L}_m,T)$ gravity, emphasizing an intriguing aspect of their structure.

\subsection{Anisotropy parameter and its influence on wormhole geometry}
 Pressure anisotropy greatly affects the geometry and stability of wormholes, playing an essential role in whether they can be traversed. To explore the different attractive and repulsive geometric shapes of wormholes, Cattoen et al.~\cite{Cattoen:2005he} and Lobo et al.~\cite{Lobo:2012qq} introduced a dimensionless anisotropy parameter. This parameter is defined as follows:
\begin{widetext}
\begin{small}
\begin{align}
 \Delta=\frac{P_t-P_r}{\rho} &= -\frac{1}{2 (\alpha+3)\Lambda r^3 \rho_{a} \left(-\delta_{c} \left(\frac{r}{r_{Sc}}\right)^{\alpha}+\left(\frac{r}{r_{Sv}}\right)^{\beta}+\delta_{c}+1\right)}\Biggl[3\Lambda \delta_{c} r^3 \rho_{a} \left(\frac{r}{r_{Sc}}\right)^{\alpha} \left(\frac{r}{r_{Sv}}\right)^{\beta} \, _2F_1\left(1,\frac{\alpha+3}{\beta};\frac{\alpha+\beta+3}{\beta};-\left(\frac{r}{r_{Sv}}\right)^{\beta}\right)\nonumber\\&+3 \Lambda \delta_{c} r^3 \rho_{a} \left(\frac{r}{r_{Sc}}\right)^{\alpha} \, _2F_1\left(1,\frac{\alpha+3}{\beta};\frac{\alpha+\beta+3}{\beta};-\left(\frac{r}{r_{Sv}}\right)^{\beta}\right)-(\alpha+3) \Lambda \delta_{c} r^3 \rho_{a} \left(\left(\frac{r}{r_{Sv}}\right)^{\beta}+1\right) \, _2F_1\left(1,\frac{3}{\beta};\frac{\beta+3}{\beta};-\left(\frac{r}{r_{Sv}}\right)^{\beta}\right)\nonumber\\&-3\Lambda \delta_{c} r_{0}^3 \rho_{a} \left(\frac{r_{0}}{r_{Sc}}\right)^{\alpha} \left(\frac{r}{r_{Sv}}\right)^{\beta} \, _2F_1\left(1,\frac{\alpha+3}{\beta};\frac{\alpha+\beta+3}{\beta};-\left(\frac{r_{0}}{r_{Sv}}\right)^{\beta}\right)+(\alpha+3) \Lambda \delta_{c} r_{0}^3 \rho_{a} \left(\left(\frac{r}{r_{Sv}}\right)^{\beta}+1\right) \, \nonumber\\& _2F_1\left(1,\frac{3}{\beta};\frac{\beta+3}{\beta};-\left(\frac{r_{0}}{r_{Sv}}\right)^{\beta}\right)-3 \Lambda \delta_{c} r_{0}^3 \rho_{a} \left(\frac{r_{0}}{r_{Sc}}\right)^{\alpha} \,  _2F_1\left(1,\frac{\alpha+3}{\beta};\frac{\alpha+\beta+3}{\beta};-\left(\frac{r_{0}}{r_{Sv}}\right)^{\beta}\right) \nonumber\\&+\alpha \Lambda r_{0}^3 \rho_{a} \left(\frac{r}{r_{Sv}}\right)^{\beta}-3 \alpha r_{0} \left(\frac{r}{r_{Sv}}\right)^{\beta}-\alpha \Lambda \delta_{c} r^3 \rho_{a} \left(\frac{r}{r_{Sc}}\right)^{\alpha}-3 \Lambda \delta_{c} r^3 \rho_{a} \left(\frac{r}{r_{Sc}}\right)^{\alpha}+\alpha \Lambda \delta_{c} r^3 \rho_{a} \nonumber\\& +\alpha \Lambda r_{0}^3 \rho_{a}-3 \alpha r_{0}+3 \Lambda r_{0}^3 \rho_{a} \left(\frac{r}{r_{Sv}}\right)^{\beta}-9 r_{0} \left(\frac{r}{r_{Sv}}\right)^{\beta}+3 \Lambda \delta_{c} r^3 \rho_{a}+3 \Delta r_{0}^3 \rho_{a}-9 r_{0}\Biggl]
\end{align}
\end{small}
\end{widetext}
When we explore the throat of a wormhole, we often find an intriguing imbalance between two types of pressure: radial pressure $P_r$ and tangential pressure $P_t$. This imbalance is what we call anisotropy $\Delta$. If this anisotropy is positive ($\Delta > 0$), it creates a repulsive effect. This is beneficial because it reduces tidal forces, making it safer and easier for travelers to journey through the wormhole. On the other hand, if the anisotropy is negative ($\Delta < 0$), we encounter an attractive effect. This increases the gravitational pull inward, complicating the journey and potentially posing risks for those who dare to traverse it. In the right panel of Fig.~\ref{fig7}, we see that cosmic voids in wormholes show negative anisotropy ($\Delta < 0$), which creates an attractive geometry. Interestingly, when $\delta_c$ gets close to zero, a small positive anisotropy ($\Delta > 0$) appears, leading to a repulsive geometry at the throat but becoming attractive again beyond that point. Moreover, anisotropic pressure is closely tied to the violation of the NEC, which is necessary for supporting traversable wormholes with exotic matter. This interplay between different pressures is key to preventing collapse and allowing for the existence of exotic spacetime geometries, giving us important insights into realistic wormhole models in the context of $f(R,\mathcal{L}_m,T)$ gravity.
 
\section{Conclusion} \label{ch: VIII}

In conclusion, this study highlights the exciting potential of cosmic voids as a pivotal element in the exploration of new wormhole solutions within the framework of $f(R,\mathcal{L}_m,T)$ gravity. By tapping into these vast underdense regions of the universe, we've developed exact static wormhole models that provide intriguing insights into how matter and spacetime interact. Our findings suggest that cosmic voids have a significant impact on the properties of wormholes, especially in shaping their gravitational effects and minimizing violations of energy conditions. One of the standout aspects of our research is the role of the parameter $\delta_{c}$. It turns out this parameter is crucial for understanding the stability and behavior of wormholes. We found that energy density remains positive across various scenarios, indicating that our models are physically sound. Interestingly, the gravitational lensing effects we observed suggest a repulsive nature, which challenges conventional ideas about gravity and opens up new possibilities for how we think about light and gravity. Moreover, our exploration into pressures at the wormhole's throat reveals a delicate balance that is essential for maintaining stability. When the radial pressure is greater than the tangential pressure, it creates a repulsive effect, making it potentially safer for hypothetical travelers. This balance is vital for ensuring that these structures can withstand disturbances, which is key to the idea of traversable wormholes. Another fascinating finding from our research is how the exoticity factor decreases as we move away from the throat. This suggests a transition from exotic to ordinary matter, which is crucial for the stability of wormholes. This dynamic nature of matter around wormholes paves the way for understanding how these structures could realistically exist. Looking ahead, our work opens up exciting avenues for further research. Cosmic voids not only provide a unique perspective on gravitational theories but also encourage us to delve deeper into how these regions can help us explore the universe's mysteries. As we refine our models and examine the implications of these voids, we may uncover groundbreaking discoveries that could change our understanding of both wormholes and the very fabric of spacetime.

In summary, this study not only enhances our grasp of wormhole dynamics in underdense environments but also lays the groundwork for future exploration of realistic traversable wormhole models. It offers a fresh perspective on the intricate relationship between cosmic voids and gravitational theories, inviting further investigation that could lead to new insights into the cosmos.

\section*{Acknowledgments }
AE thanks the National Research Foundation of South Africa for the award of a postdoctoral fellowship. 

\section*{Conflict Of Interest statement }
The authors declare that they have no known competing financial interests or personal relationships that could have appeared to influence the work reported in this paper.

\section*{Data Availability Statement} 
This manuscript has no associated data, or the data will not be deposited. (There is no observational data related to this article. The necessary calculations and graphic discussion can be made available
on request.)

\end{document}